\documentclass[11pt]{article}
\usepackage{a4,graphicx,xspace,oupbib,amsmath,amsthm}
\usepackage{algorithm}
\usepackage{xcolor}
\usepackage[noend]{algpseudocode}

\newcommand{\be}{\begin{equation}}
\newcommand{\bel}[1]{\begin{equation}\label{#1}}
\newcommand{\ee}{\end{equation}}

\newcommand{\LR}{\text{LR}}

\newcommand{\DNAmixtures}{{\tt DNAmixtures}}

\newcommand{\KinMix}{{\tt KinMix}}

\renewcommand{\H}{\mathcal{H}}
\newcommand{\Hp}{\H_{\text{p}}}

\newcommand{\n}{\boldsymbol{n}}
\newcommand{\z}{\boldsymbol{z}}
\newcommand{\ub}[1]{\underline{\boldsymbol{#1}}}

\newcommand{\eg}{{\em e.g.\/}\xspace}
\newcommand{\etc}{{\em etc.\/}\xspace}

\begin{document}
\title{{\sc Inference about complex relationships using peak height data from {DNA} mixtures}}
\author{
Peter J. Green\thanks {School of Mathematics, University of
Bristol, Bristol BS8 1TW, UK.
\newline \hspace*{5mm} Email: {\tt P.J.Green@bristol.ac.uk}}\\
University of Bristol, and\\
University of Technology, Sydney.\\
\and Julia Mortera\thanks {Universit\`a Roma Tre, Italy.
\newline \hspace*{5mm} Email: {\tt julia.mortera@uniroma3.it}}\\
Universit\`a Roma Tre. \\
}
\date{\today}
\maketitle

\begin{abstract}
In both criminal cases and civil cases there is an increasing demand
for the analysis of DNA mixtures involving relationships. The goal
might be, for example, to identify the contributors to a DNA mixture where the donors may
be related, or to infer the relationship between individuals based on a mixture. 

This paper introduces an approach to modelling and computation for DNA mixtures
involving contributors with arbitrarily complex relationships. It builds on an
extension of Jacquard's condensed coefficients of identity, to specify and compute with joint
relationships, not only pairwise ones, including the possibility of inbreeding.

The methodology developed is applied to two casework examples involving a missing person,
and simulation studies of performance, in which the ability of the methodology
to recover complex relationship information from synthetic data with known `true' 
family structure is examined.

The methods used to analyse the examples are implemented in the new
\verb+KinMix+ \textbf{R} package, that extends the \verb+DNAmixtures+ package to allow for
modelling DNA mixtures with related contributors. 
%\verb+KinMix+ inherits from \verb+DNAmixtures+ the capacity to deal with mixtures with many contributors, in a time- and space-efficient way.

\hspace{5mm}

\noindent {\small {\em Some key words:} Bayesian networks, coefficients of identity, criminal identification, disputed paternity, 
DNA mixtures, identity by descent, inbreeding, kinship, uncertainty in allele frequencies.}

\end{abstract}

\section{Introduction}

This article is concerned with probabilistic genotyping methods for DNA mixtures based on unlinked autosomal short tandem repeat (STR) markers, under hypotheses about biological relationships involving contributors to the mixture. For the first time, answers to important questions about complex relationships can be delivered by a fully probabilistic approach.

\subsection{Genetic background}

STR markers are the mainstay of DNA profiling systems used in forensic science laboratories world-wide. 
To create the data used in modelling and analysis, the DNA in a biological specimen is extracted, and through a process of polymerase chain reaction (PCR) amplified to create an electropherogram (EPG), a continuous trace that is digitised for analysis. Identifiable segments of this trace correspond to different known loci, called markers, on the genome. The markers used are chosen to lie far apart in the genome, usually on different chromosomes, so are unlinked, that is statistically independent. Each segment consists of peaks aligned to a known grid, one peak for each possible allele, or genetic variant, at that locus; for STR markers, these alleles take (generally, integer) values giving the repeat count of a short sequence in the genome. The heights of the peaks measure the abundance of that allele at that locus, and collectively carry a great deal of information about the genetic make-up of the original specimen, but are subject to degradation consisting of both random noise and artefacts introduced in the PCR process, for which a statistical model must be constructed. 

The usual methodology uses only autosomal loci, that is, markers on the chromosomes that are not sex-linked, and so involve a contribution from each parent. The individual contributions we call genes, and the unordered pair of contributions the genotype. We call the collection of genotypes across all markers the genotype profile of the individual; note that in a statistical language, we regard genes and genotypes as random variables, while alleles are the possible values of a gene.

When a biological specimen is taken from one individual, and handled and processed under ideal conditions, an EPG represents a noisy version of the genotype profile of that one individual; in this case we can talk about genotyping or `typing'. In other situations, either through the circumstances in which the sample is obtained (for example, in rape, or other violent assault) or through contamination, the DNA quantified by the EPG comes from two or more individuals; this is a DNA mixture. In both criminal and civil cases, even for simple tasks such as criminal identification or paternity testing, we need methodology for assessing hypotheses about the genetic origin of the DNA based on mixture data. The focus of this paper is to derive and demonstrate principled and practical methods for such tasks, in cases where there are biological relationships involving the contributors. We concentrate on so-called continuous methods using peak height information, as opposed to binary or semi-continuous methods.

More details of the genetic background to the use of STR markers and DNA mixtures can be found in Section 1 of \textcite{cowell:etal:15}, in this Journal, and also in the recent review article by \textcite{mortera-arsia}. 

\subsection{Modelling and inferential approach}

Both the statistical model we use for EPG peak heights, and exact computationally-efficient methods for inference about the genotype profiles of contributors, can be found in \textcite{cowell:etal:15}. The present article extends the principles and practice of this work to allow for relationships among the contributors and/or to make inferences about such relationships. We stress here that we are concerned with \emph{relationships}, that is specific close familial associations such as between parent and child, or an individual and her maternal grandparents, not \emph{relatedness}, that is an ambient genetic correlation induced among all members of a population or subgroup by virtue of shared ancestry, typically quantified by a coancestry coefficient.

There has been very little previous work on models, methodology and software for DNA mixtures that takes account of relationships with and among the contributors. The only examples we are aware of are our earlier work on paternity analysis in \textcite{green:mortera:17}, the precursor to the present paper, and \textcite{hernandis2019relmix}. Our work models peak heights explicitly, while \textcite{hernandis2019relmix} does not use peak height information; these two approaches underpin the {\tt KinMix} and {\tt relMix} software packages respectively. There are several other packages dealing with mixtures in the presence of relatedness; some of these are listed in the Supplementary information, section 3, but we will not be concerned further with such models here; they are not capable of modelling close familial relationships accurately.

It is convenient and natural when modelling peak height data probabilistically to use a hierarchical formulation, with two main layers: the genotype profiles $\n$ of the contributors, and the peak heights $\z$ recorded in the electropherogram; the models typically then consist of two components: 
\begin{enumerate}
\item $p(\n)$ -- the joint distribution for $\n$ -- parameterised by population allele frequencies, hypotheses about the contributors, etc., and 
\item $p(\z|\n)$ -- the conditional distribution for $\z$ given $\n$ -- with parameters identifying the peak height model and the proportions of DNA from each of the contributors contained in the mixture.
\end{enumerate}

Inference based on DNA mixtures usually focusses on the comparison between two hypotheses $\Hp$ and $\H_0$ concerning the constitution of the mixture, quantified by the likelihood ratio \LR\ for $\Hp$ versus $\H_0$. In this article, these hypotheses will be about (arbitrarily complex) relationships between mixture contributors, and between contributors and other typed individuals. Simple examples of relationship tests we can construct are
\begin{itemize}
\item a paternity test given a child's genotype, where $\Hp$ and $\H_0$ respectively state that the putative father is a contributor to the mixture, or that no contributor to the mixture is related to the child, and  
\item a test for whether contributors to a mixture, perhaps found at a crime scene, are related in a particular way ($\Hp$) or not at all ($\H_0$).
\end{itemize}

Although above and throughout the paper, we use the words `test' and `hypothesis', etc., that are familiar from statistical hypothesis testing, we should make clear that this is not what we are doing. Rather, we are contrasting alternative explanations of the evidence, by means of $\log_{10} \LR$ values. These `weights of evidence' are to be interpreted in the applicable judicial context, where they will be combined with other evidence, not necessarily quantifiable numerically, rather than referred to a null distribution to yield for example p-values. The $\log_{10} \LR$ scale for weights of evidence is standard in forensic statistics \cite{balding:05}, but in general usage can be traced back to Turing; see \textcite{good}. The scale is natural and interpretable: for example a $\log_{10} \LR$ of 6 means that the evidence is $10^6=1,000,000$ times more likely under one hypothesis than another.

Throughout we use the probabilistic and computational formulation for DNA mixtures of \textcite{cowell:etal:15}, and the software implementation of this in the {\tt DNAmixtures} {\bf R} package of \textcite{graversen:package:13}; the model emulates the PCR process described above, and recognises artefacts including stutter, drop-out, drop-in and silent alleles. Our new model extensions are largely aimed at modelling the genotype profile distribution $p(\n)$ to express complex relationships; the methods are implemented in an {\bf R} package {\tt KinMix} \cite{kinmix} that supplements {\tt DNAmixtures}. Early ideas in this direction can be found in \textcite{green:mortera:17}. Although implementation is restricted to this model and this computational environment, the ideas are quite general and could be adapted to other probabilistic genotyping systems, and to other peak height models.

\subsection{Contribution of this paper, and its structure }

The novel methodological contribution of this paper is to show both how recent unpublished work on extending coefficients of identity to more than two individuals provides a formalism for specifying complex family relationships, and how this can be used to automatically construct Bayesian network (BN) algorithms for computing inferences for models for DNA mixtures involving related individuals, including likelihood ratios for comparing hypotheses about relationships.

The paper is organised as follows. In Section 2, we present a formulation and notation for expressing quantitatively the simultaneous relationships among an arbitrary set of individuals, specified through their joint pedigree. This notation generalises long-standing coefficients of relationship applicable only to pairs of individuals, and is useful both for specifying relationships precisely and computing with them. The formulation is completely general, its applicability is not confined to STR markers and DNA mixtures, but encodes relationships for any quantitative genetic analysis. In Section 3, we use this encoding to characterise the joint distribution of genotypes in a family, and thereby develop efficient computational inference schemes for DNA mixtures, using Bayesian network (BN) methods. Section 4 is devoted to BN computations for mixtures where there is an `ambient' level of relatedness across the whole population, rather than specific close relationships, or where allele frequencies are not exactly known.  In section 5 we discuss the setting of parameters for likelihood calculations in our model, and in section 6 we introduce the {\bf R} package \KinMix\ implementing all of the methods in this paper. Section 7 and 8 illustrate the methodology with some simulated scenarios, and several real case studies, including some comparative timings highlighting the fast computation times that are achieved.  Finally, we discuss extensions such as allowing for mutation.

\section{Encoding relationships via IBD pattern distributions}

What are biological relationships, and how can they be encoded? An essential preliminary to modelling and analysis of DNA mixtures with related contributors is a compact and precise representation of joint relationships among a set of arbitrarily related individuals. Developing and illuminating such a representation is the focus of this section, which is of universal applicability, not confined to mixture analysis.

Under our simplified genetic model of unlinked autosomal STR markers, the sole source of relationship between individuals is \emph{identity by descent} (IBD). This is the phenomenon that two genes may be identical because they are copies of the same ancestor gene, rather
than being independent draws from the ‘gene pool’, so that the genotypes of two or more related actors will be positively associated. 

It is important to distinguish IBD from \emph{identity by state} (IBS), which includes the possibility that independent draws from the gene pool have the same value by chance. For example, consider a mother--daughter pair, and a single STR marker where the mother's genotype is say ($a$,$b$), and the child's is ($c$,$d$), where in both cases we have ordered the genes in the genotype as (maternal gene, paternal gene). Then by Mendel's first law, $c$ is equal to $a$ or $b$ with equal probability, while $d$ is contributed by the father. The gene $c$ is identical by descent to either $a$ or $b$ respectively, but unless the mother's and father's genes are all distinct in value, this is not the only way that any of ($a$,$b$,$c$,$d$) can be equal in value (identical by state). For example, $a$ and $b$ could by chance be equal (i.e. the mother is homozygous), or $a$ and $d$ could be equal, or indeed both.

For relationships other than parent--child, and for relationships among more than two individuals, we need a way to encode the implications of Mendel's first law for the joint distribution of all the individuals' genotypes; that is the objective of this section.

A pedigree is a graphical or tabular depiction of the identities of all individuals under consideration, mapping to the identities of their parents, and is a useful compact and precise vehicle for specifying joint relationships. In all of the pedigrees we use and display, either both or neither of the parents are included; those with no parents in the pedigree are called founding individuals, or founders, and their genes are founding genes.  Given a pedigree, IBD is determined by the meioses generating the genes of each child given those of its parents, and any inbreeding within or between founding individuals: the actual allelic values are not relevant to this. For two individuals, Table 1 of \textcite{thompson:genetics} lays out all possible patterns; this table is adapted as our Table \ref{tab:thompson} below. An explanation of the IBD states is given in Section \ref{sec:simple}.
Thompson credits this formulation to \textcite{nadot:vaysseix}, although they do not use a tabular representation. 

\begin{table}[ht]
\caption{The IBD states among the four genes of two individuals $B_1$ and $B_2$, adapted from Table 1 of \textcite{thompson:genetics}.
The two genes of individual $B_1$ are denoted $a$, $b$, and those of $B_2$ are $c$ and $d$.
The IBD state is defined by the labelling developed by \textcite{nadot:vaysseix}, and further explained in Section \ref{sec:simple}.
For example, the fifth line refers to the possibility that the two genes for $B_1$ are identical by descent, while those for $B_2$ are not, and are not identical by descent with those for $B_1$: that pattern of identity is shown equivalently as both an IBD state, a string of labels $(1,1,2,3)$ where labels are equal if and only if there is identity by descent, and as a partition, $(a,b)(c)(d)$.
The states can be grouped into subsets of genotypically equivalent states, indicated by the horizontal lines; the total probability $\Delta_i$ of the subset is given on the first row of each. For example, $\Delta_3$ is the combined probability of states $(1,1,1,2)$ and $(1,1,2,1)$.
\label{tab:thompson}
}
\centering
	\vspace{2mm}
    \begin{tabular}{rrrrccc}
\hline
 \multicolumn{4}{c}{IBD state} &  & &\\
 \multicolumn{2}{c}{$B_1$} &\multicolumn{2}{c}{$B_2$} &  &   \multicolumn{2}{c}{Probability} \\
$a$ & $b$ & $c$ & $d$ & Partition & Jacquard & \boldmath$\kappa$ \\
        \hline
         1 & 1 & 1 & 1&$(a,b,c,d)$ & $\Delta_1$ & --\\
				\hline
        1 & 1 & 2 & 2&$(a,b)(c,d)$ & $\Delta_2$ & --\\
				\hline
        1 & 1 & 1 & 2&$(a,b,c)(d)$ & $\Delta_3$ & --\\
        1 & 1 & 2 & 1&$(a,b,d)(c)$ &  & --\\
				\hline
         1 & 1 & 2 &3&$(a,b)(c)(d)$ & $\Delta_4$ & --\\
				\hline
        1 & 2 & 1 & 1&$(a,c,d)(b)$ & $\Delta_5$ & --\\
        1 & 2 & 2 & 2&$(a)(b,c,d)$ &  & --\\
				\hline
        1 & 2 &3 & 3 &$(a)(b)(c,d)$ & $\Delta_6$ & --\\
				\hline
         1 &2 & 1 & 2&$(a,c)(b,d)$ & $\Delta_7$ & $\kappa_2$\\
        1 & 2 & 2 & 1&$(a,d)(b,c)$ &  & --\\
				\hline
        1 & 2 & 1 & 3&$(a,c)(b)(d)$ & $\Delta_8$ & $\kappa_1$\\
        1 & 2 & 3 & 1&$(a,d)(b)(c)$ &  & --\\
        1 & 2 & 2 & 3&$(a)(b,c)(d)$ &  & --\\
        1 & 2 & 3 & 2&$(a)(b,d)(c)$ &  & --\\
				\hline
        1 & 2 & 3 & 4&$(a)(b)(c)(d)$ & $\Delta_9$ & $\kappa_0$\\
    \hline
\end{tabular}
\end{table} 
								
\subsection{Coefficient of identity by descent}
\label{sec:coefIBD}

Two-person relationships are compactly summarised in numerical form using the \emph{coefficients of identity by descent} ($\delta_i$) and \emph{condensed coefficients of identity by descent} ($\Delta_i$) of \textcite{jacquard} (chapter 6), which are probabilities of particular patterns of identity by descent. The $\delta_i$ are the probabilities for the 15 individual rows of Table 1, the $\Delta_i$ those of the 9 subsets of genotypically equivalent states, where we do not keep track of which parent donates which allele; in each case the numbering of the coefficients follows that of the respective original authors, and conforms with the natural ordering in the Table. It is these condensed coefficients of identity that we use to characterise and quantify relationships among related individuals, including mixture contributors. Where inbreeding is ruled out, $\Delta_i=0$ for $i=1,2,\ldots,6$, and we need only the $\kappa$ coefficients of \textcite{cotterman}, which
have an explicit interpretation: $\kappa_0\equiv\Delta_9,\kappa_1\equiv\Delta_8,\kappa_2\equiv\Delta_7$ are the probabilities that the two individuals share 0, 1 or 2 alleles by descent. As examples,
\begin{enumerate}
\item a parent and child have a relationship summarised by $\kappa_0=0, \kappa_1=1, \kappa_2=0$ (since a parent always contributes exactly one of its alleles to his or her child), 
\item two half-siblings (with unrelated parents) have $\kappa_0=0.5, \kappa_1=0.5, \kappa_2=0$ (since there is a 50--50 chance that the common parent contributes the same allele to each of its children), while 
\item two children from an incestuous brother-sister mating are captured by $\Delta=(0.06250,0.03125,$ $0.12500,0.03125,0.12500,0.03125,0.21875,0.31250,0.06250)$,  a more complicated example demonstrating the need to automate the calculation of these coefficients. 
\end{enumerate}
The $\kappa$ and $\Delta$ coefficients can be calculated from a pedigree by simple recursive calculations down the pedigree. In {\bf R}, these are performed by the functions \verb+kappaIBD+ and \verb+condensedIdentity+, respectively, in the package \verb+ribd+ \cite{ribd}, part of the \verb+pedtools+ family of packages created by \textcite{pedtools}.

The coefficients of identity, $\kappa_i$, $\delta_i$ or $\Delta_i$, should not be confused with a different measure of relatedness, the coefficient of coancestry or
kinship coefficient, often represented by $\theta$. This is the probability that randomly chosen alleles from each of the two individuals are identical by descent. By elementary probability calculations, we find that $\theta=\Delta_1+(\Delta_3+\Delta_5+\Delta_7)/2+\Delta_8/4$, or simply $\kappa_2/2+\kappa_1/4$ in the absence of inbreeding. This carries less information, for example, your coefficient of ancestry with your mother and sister are both $\theta=1/4$, although you are differently related to these two, as shown by the $\kappa$s. Similarly, you have the same $\theta=1/8$ with a double first cousin and an uncle.

For more than 2 individuals, \textcite{Thompson74} seems to have been first to provide a general framework for gene identity given multiple relationships. She provides
a rigorous algebraic formalism, with particular attention to enumerating the intrinsic symmetries in the problem, and counts the numbers of possible relationships, which increase very rapidly. For example, for as few as 4 individuals, there are already 712 possible (genotypically equivalent) IBD states, reducing to 139 if inbreeding is ruled out. She does not provide a notation for the probabilities of these states. We could extend the $\Delta$ notation to more than two individuals, using as subscript the IBD pattern, for example writing $\Delta_{1,2,1,3}$ in place of $\Delta_8$, but this seems cumbersome, especially for larger numbers of relatives.
In typical pedigrees, only a very small fraction of these states have positive probability, so the vast majority of condensed coefficients of identity are 0. Instead, we use what amounts to a sparse representation of such vectors of coefficients, namely a listing of which coefficients are non-zero, and their probability values; there are examples in Tables \ref{IBDpatt} and \ref{tab:3cousins}, where we use an arbitrary member of each equivalence class as a representative. We call this representative an IBD pattern and the whole table the \emph{IBD pattern distribution}. 
In recent unpublished collaborative work, the present first author and Magnus Dehli Vigeland jointly devised an efficient algorithm for computing such tables, also known as `multi-person condensed coefficients of identity', from the pedigree; this is implemented in the function {\tt pedigreeIBD} in the {\bf R} package {\tt KinMix} and the function {\tt multiPersonIBD} in the {\bf R} package {\tt ribd}.

We can display each IBD pattern in various ways, following the authors cited above. One is by a vector of integer labels, of length twice the number of individuals, $n$, say, the pair in entries $(2i-1,2i)$ representing the genotype of the corresponding individual $i=1,2,\ldots,n$, as in the first columns of Table \ref{tab:thompson}. The numerical value of the labels is irrelevant, all that matters is whether two labels are the same or different, so the vector denotes a partition of the $2n$ genes according to which are identical by descent. Since we are only concerned with unordered pairs of genes, the interpretation of the pattern is unchanged if elements $(2i-1,2i)$ are exchanged, and also, of course, unchanged by any 1--1 relabelling. Diagrammatically, the pattern can be displayed as a graph with $2n$ vertices laid out in a $n\times 2$ rectangular array, and vertices connected by an arc if the corresponding genes are identical by descent. Both representations of IBD patterns were used (with $n=2$) by \textcite{jacquard} (chapter 6). Figure \ref{fig:3cousins} and Table \ref{tab:3cousins} show two examples of IBD pattern distributions displayed in each of these ways.

As a foretaste of what can be done with the $\kappa$ and $\Delta$ coefficients, or their generalisation the IBD pattern distribution, we remark that, together with the population alleles frequencies, they determine the joint distribution of the genotypes for any set of individuals; as a simple example, if $a$, $b$ and $c$ are distinct alleles, with population frequencies $q_a,q_b,q_c$, and in the absence of inbreeding, the probability that two individuals have genotypes $(a,b)$ and $(a,c)$ respectively is simply $\kappa_0(4q_a^2q_bq_c)+\kappa_1(q_aq_bq_c)$, when their relationship is summarised by $\boldmath\kappa=(\kappa_0,\kappa_1,\kappa_2)$. This can be verified directly algebraically with care, but in Section \ref{sec:specific} we generalise this calculation and use it in modelling and analysing DNA mixtures.

\subsection{IBD pattern distribution for a simple pedigree}
\label{sec:simple}

As an illustration, consider a simple `triple' of father F, mother M and child C, with the two parents unrelated. If we label the father's genes by (1,2) and those of the mother by (3,4), then the child will have one gene that is either 1 or 2, and another gene that is 3 or 4; thus its genotype is (1,3), (2,3), (1,4) or (3,4) with equal probability. In tabular form, thus, we could write this family's genetic structure at any single autosomal locus in a table as in panel (a) of Table \ref{IBDpatt}, where we have labelled the columns with the individual identities, and the rows with the corresponding probabilities.

\begin{table}[ht]
\caption{IBD pattern distributions for a Father/Mother/Child triple, F, M, C.}\label{IBDpatt}
\begin{center}
(a) Distinguishing maternal and paternal genes\\[3mm]
\begin{tabular}{c|cccccc}
\hline
pr & \multicolumn{2}{c}{F} & \multicolumn{2}{c}{M} & \multicolumn{2}{c}{C} \\
\hline
0.25 & 1 & 2 & 3 & 4 & 1 & 3 \\
0.25 & 1 & 2 & 3 & 4 & 2 & 3 \\
0.25 & 1 & 2 & 3 & 4 & 1 & 4 \\
0.25 & 1 & 2 & 3 & 4 & 2 & 4 \\
\hline
\end{tabular}
\end{center}
\begin{center}
(b) Condensed form: Not distinguishing maternal and paternal genes\\[3mm]
\begin{tabular}{c|cccccc}
\hline
pr & \multicolumn{2}{c}{F} & \multicolumn{2}{c}{M} & \multicolumn{2}{c}{C} \\
\hline
1 & 1 & 2 & 3 & 4 & 1 & 3  \\
\hline
\end{tabular}
\end{center}
\begin{center}
(c) Extending the family to include the paternal grandfather\\[3mm]
\begin{tabular}{c|cccccccc}
\hline
pr & \multicolumn{2}{c}{F} & \multicolumn{2}{c}{M} & \multicolumn{2}{c}{C} & \multicolumn{2}{c}{GF}\\
\hline
0.5 & 1 & 2 & 3 & 4 & 1 & 3 & 1 & 5\\
0.5 & 1 & 2 & 3 & 4 & 1 & 3 & 2 & 5\\
\hline
\end{tabular}
\end{center}
\end{table}

Note that without changing the meaning, we can arbitrarily permute the actual labels, separately in each row, so the first row of Table \ref{IBDpatt}(a) could have been (2,4,1,3,2,1); the purpose of the labels is solely to indicate which genes are identical (by descent) and which different.
Since a genotype is an \emph{unordered} pair of genes, the interpretation of the table is also unchanged
if any of the individual pairs are transposed, so the first row could equivalently be written (2,1,4,3,1,3), for example, and in many other ways. Combining these two rules, and aggregating the probabilities of identical rows,
further economy of notation is possible: for example we could simply use the table in panel (b) of Table \ref{IBDpatt},
to represent the same family, saving space and computer time. Effectively the labels 1 and 3 are then being used for the child's paternal and maternal genes respectively. 

The example can be extended, by, for example, including also the Father's father, GF. There are two equally likely possibilities: the gene inherited by Father from his father might be that labelled 1 or 2. So the relationships between the 4 individuals can now be represented by panel (c) of Table \ref{IBDpatt}. 

\subsection{Pairwise relationships do not determine joint relationships}
\label{sec:joint}

\begin{figure}[htbp]
\centering
  \resizebox{0.8\textwidth}{!}{\includegraphics{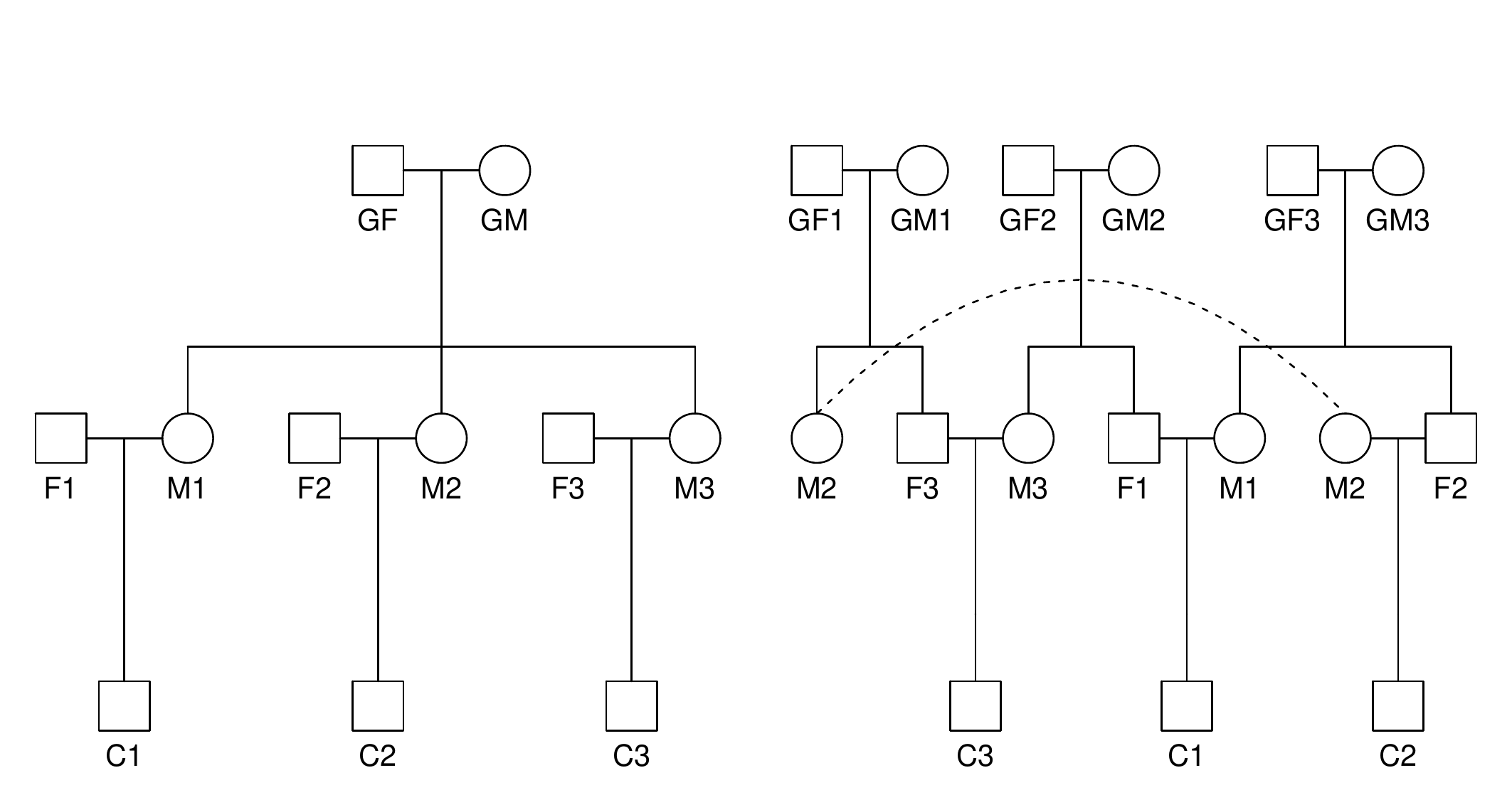}}
\caption{Pedigrees for the two 3-cousins scenarios: (left) star, (right) cyclic; one individual appears twice, to reduce line-crossing -- the dotted curves link the replicate symbols.}\label{fig:3cousins-ped}
\end{figure}

To demonstrate that pairwise relationships do not determine a full description of relatedness among more than two individuals, even in the absence of inbreeding, we present the simplest example. Consider two scenarios in which among three individuals, each pair are full cousins, that is have $(\kappa_0,\kappa_1,\kappa_2)=(0.75,0.25,0)$. This can arise in a `star' arrangement, where the three have mothers who are full siblings, but unrelated fathers (or vice-versa, of course). In a `cyclic' arrangement, each pair of cousins have between them parents of the opposite sex who are siblings, with the other parents unrelated. The two pedigrees are displayed in Figure \ref{fig:3cousins-ped}.

\begin{figure}[htbp]
\centering
\begin{tabular}{c|c}
  \begin{minipage}{0.45\textwidth}
  \centering
  \resizebox{8cm}{!}{\includegraphics{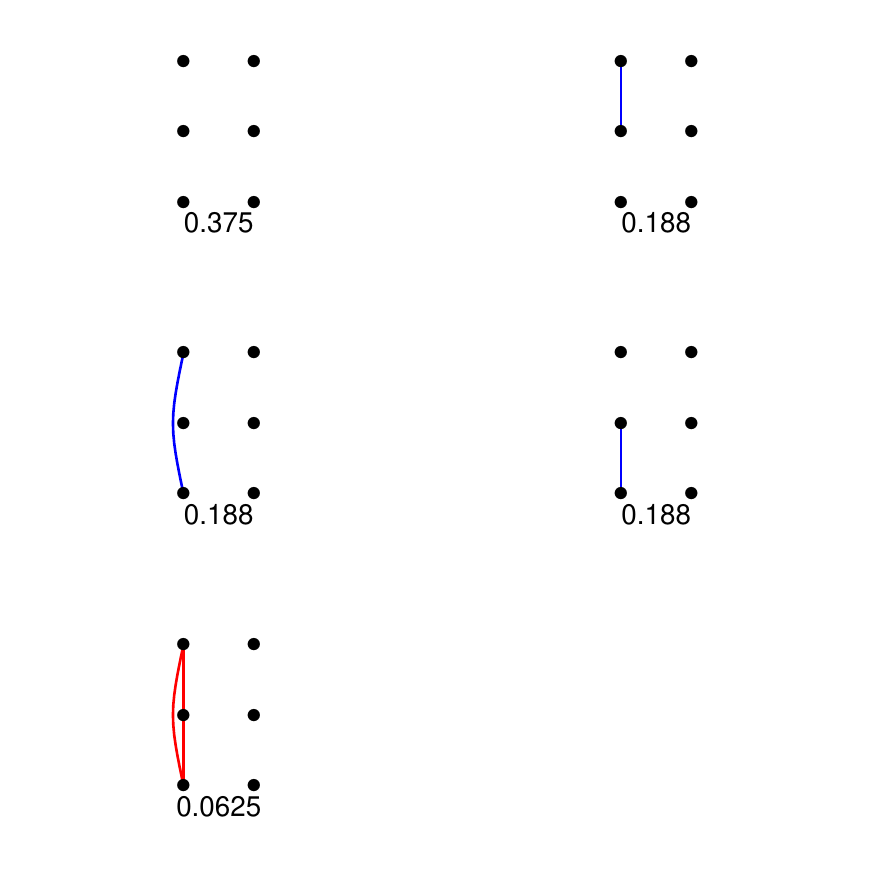}}
  \end{minipage}
	&
  \begin{minipage}{0.45\textwidth}
  \centering
  \resizebox{8cm}{!}{\includegraphics{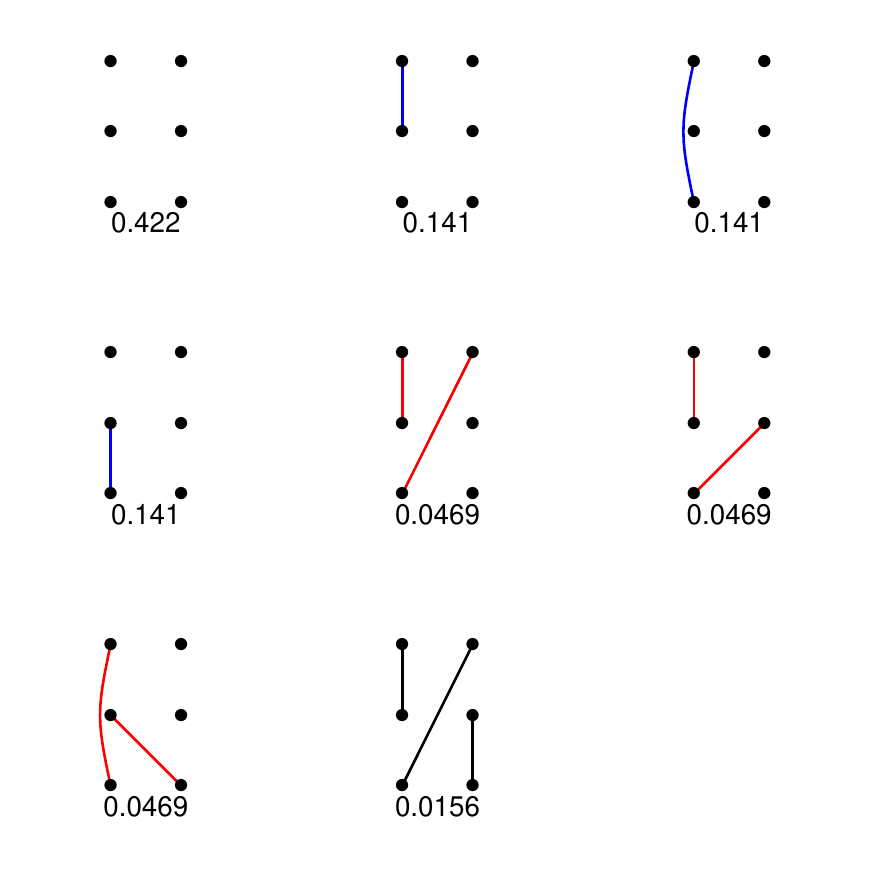}}
  \end{minipage}
\end{tabular}
\caption{ IBD pattern distributions for two scenarios of 3 pairwise cousins; (left) star, (right) cyclic arrangements, represented in the style introduced by \textcite{jacquard} and described in the text. In each individual diagram, from top to bottom, the individuals are those labelled C1, C2 and C3, respectively, in Figure \ref{fig:3cousins-ped}, and the corresponding probability is printed below the diagram. As an example, the final diagram in the right hand panel shows the case where each pair of C1, C2 and C3 have a gene identical by descent, which is represented by the IBD pattern (1,2,1,3,2,3), which is how it is labelled in the final row of the right hand panel of Table \ref{tab:3cousins}, where it is assigned the same probability, 0.0156.}\label{fig:3cousins}
\end{figure}

\begin{table}[ht]
\caption{IBD pattern distributions for two scenarios of 3 pairwise cousins; (left) star, (right) cyclic arrangements.}\label{tab:3cousins}
\begin{center}
\begin{tabular}{c|cccccc||c|cccccc}
\hline
pr & \multicolumn{2}{c}{C1} & \multicolumn{2}{c}{C2} & \multicolumn{2}{c||}{C3} & pr & \multicolumn{2}{c}{C1} & \multicolumn{2}{c}{C2} & \multicolumn{2}{c}{C3} \\
\hline
0.3750 & 1 & 2 & 3 & 4 & 5 & 6 & 0.4219 & 1 & 2 & 3 & 4 & 5 & 6 \\
0.1875 & 1 & 2 & 1 & 3 & 4 & 5 & 0.1406 & 1 & 2 & 1 & 3 & 4 & 5 \\
0.1875 & 1 & 2 & 3 & 4 & 1 & 5 & 0.1406 & 1 & 2 & 3 & 4 & 1 & 5 \\
0.1875 & 1 & 2 & 3 & 4 & 3 & 5 & 0.1406 & 1 & 2 & 3 & 4 & 3 & 5 \\
0.0625 & 1 & 2 & 1 & 3 & 1 & 4 & 0.0469 & 1 & 2 & 1 & 3 & 2 & 4 \\
&&&&&&& 0.0469 & 1 & 2 & 1 & 3 & 3 & 4 \\
&&&&&&& 0.0469 & 1 & 2 & 3 & 4 & 1 & 3 \\
&&&&&&& 0.0156 & 1 & 2 & 1 & 3 & 2 & 3 \\
\hline
\end{tabular}
\end{center}
\end{table}

The respective IBD pattern distributions are shown in Table \ref{tab:3cousins} and visualised in Figure \ref{fig:3cousins}; the formats of each are described at the end of Section \ref{sec:coefIBD}. It is very clear from these displays that these two families have different overall relatedness, for example in the star arrangement, there is probability 0.0625 that the three cousins have a common allele by IBD (as seen in the left-hand panels of Table \ref{tab:3cousins} and Figure \ref {fig:3cousins}), while this is impossible in the cyclic arrangement.
The fact that $(\kappa_0,\kappa_1,\kappa_2)=(0.75,0.25,0)$ for each pair of cousins, for both pedigrees, can be verified from the Table. Taking C1 and C2 for example, in none of the rows in either half of the table, are the labels for C1 and C1 both the same, so $\kappa_2=0$. Meanwhile, the 4 labels are all distinct in rows 1, 3, and 4 in the left panel, and 1, 3, 4 and 7 in the right panel, so we can calculate $\kappa_0$ as $0.3750+0.1875+0.1875=0.4219+0.1406+0.1406+0.0469=0.75$.

In Section \ref{sec:simulations}, we include a simulation experiment demonstrating the extent to which these pedigrees can be distinguished from DNA mixtures of STR markers with these family members as contributors. 

We noted in Section \ref{sec:coefIBD} that pairwise relationships are not correctly quantified by coefficients of ancestry $\theta$, so \emph{a forteriori} neither are joint relationships between more than 2 individuals.

\subsection{IBD pattern distribution in the Iulius-Claudius pedigree}
\label{sec:caesars}

\begin{figure}[htbp]
	\begin{center}
		\includegraphics
		%    [bbllx=47,bblly=29,bburx=548,bbury=743,scale=0.9,clip]
		[width=.95\textwidth] {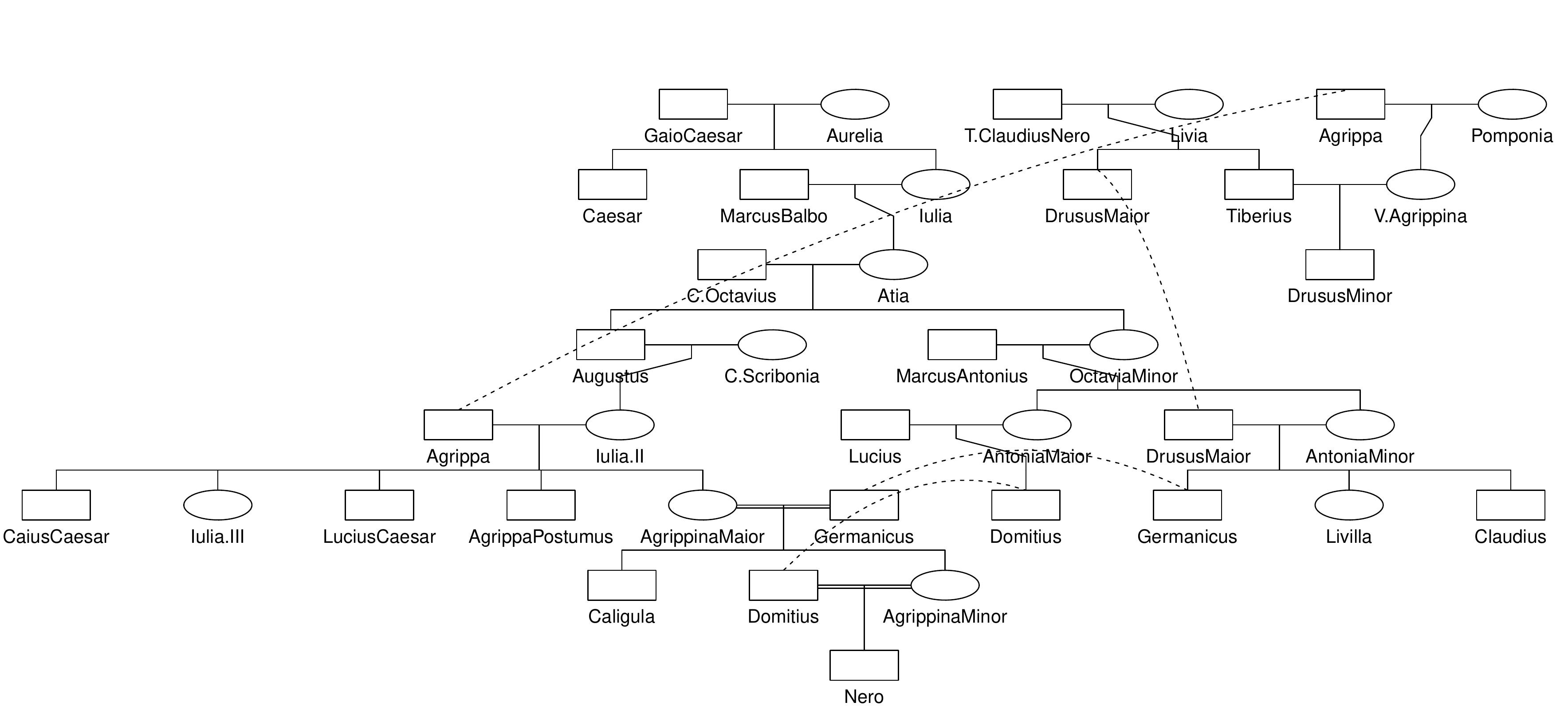}
		\caption{Iulius-Claudius family tree}
		\label{fig:Caesar}
	\end{center}
\end{figure}

As a demonstration that the IBD pattern distribution among a handful of individuals can be readily computed even when the pedigree describing their relationships consists of as many as 35 individuals, and inbreeding is present, we briefly consider the Iulius-Claudius dynasty. This was the first Roman imperial dynasty, consisting of the first five emperors -- Augustus, Tiberius, Caligula, Claudius, and Nero -- and  the family to which they belonged. They ruled the Roman Empire from its formation under Augustus in 27 BCE until 68 CE, when the last of the line, Nero, committed suicide. The name Iulius-Claudius  dynasty refers to the two main branches of the imperial family: the gens Iulia  and the gens Claudia. Figure \ref{fig:Caesar} presents the Iulius-Claudius pedigree. Some of the names have been abbreviated.

\begin{figure}[htbp]
\centering
\begin{tabular}{c|c}
  \begin{minipage}{0.45\textwidth}
  \centering
  \resizebox{8cm}{!}{\includegraphics{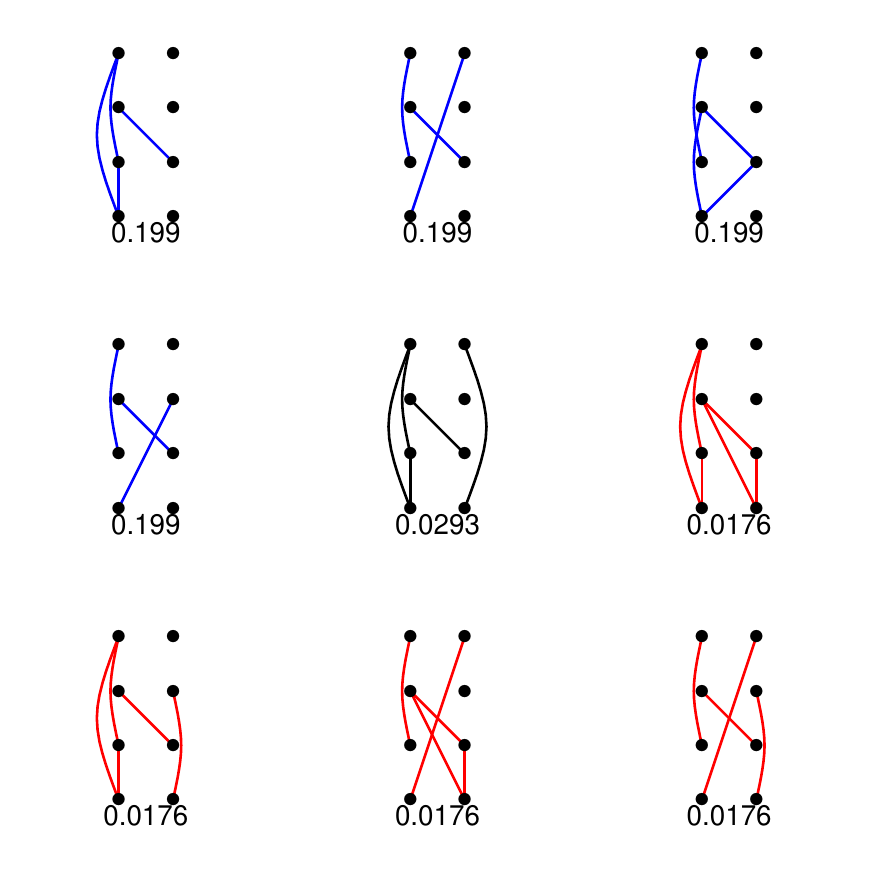}}
  \end{minipage}
	&
  \begin{minipage}{0.45\textwidth}
  \centering
  \resizebox{8cm}{!}{\includegraphics{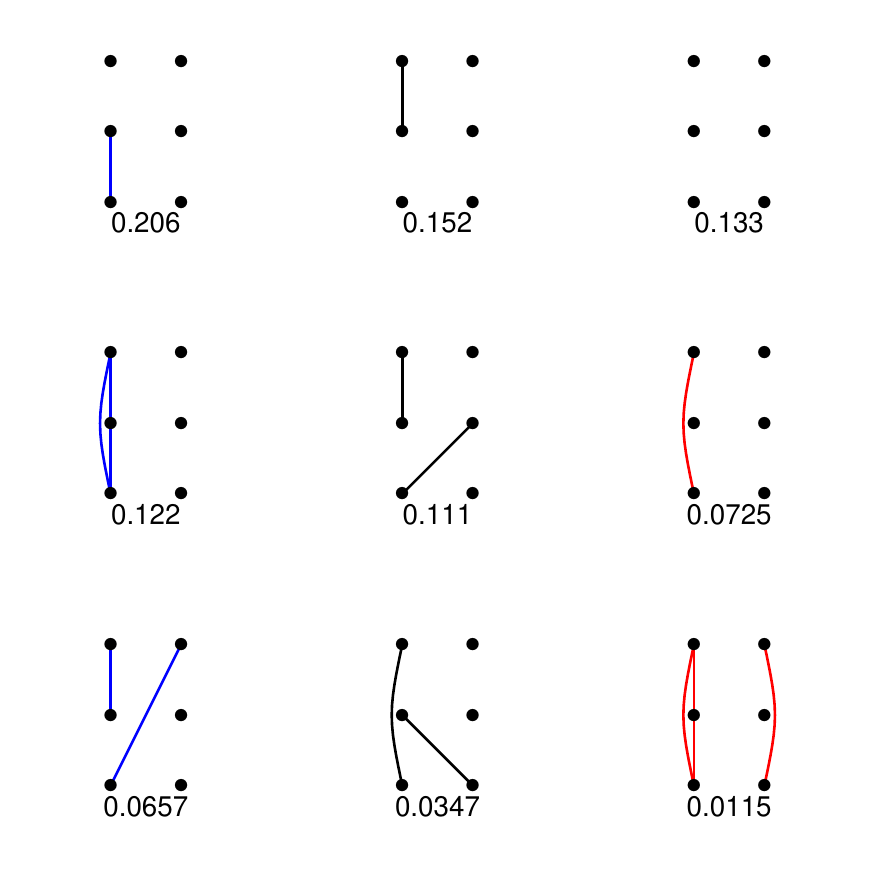}}
  \end{minipage}
\end{tabular}
\caption{(left) Most probable patterns in the IBD pattern distributions for (respectively from the top) Germanicus, his wife Agrippina Maior and their descendants
		the Emperors Caligula and Nero, (right) the same for the three Emperors Claudius, Caligula and Nero.}		\label{fig:4emperors}
\end{figure}

As highlighted by the double horizontal bars in Figure \ref{fig:Caesar}, there are two inbred marriages within the pedigree, the first, between Germanicus and Agrippina Maior, leading to the other. The two emperors Caligula and Nero are descendants of this couple. We therefore look at the IBD pattern distribution among Germanicus, Agrippina Maior, Caligula and Nero, and also that between the three emperors Claudius, Caligula and Nero. Figure \ref{fig:4emperors} shows excerpts (the most probable 9 patterns) of the IBD pattern distribution for these two subsets of individuals, respectively. As one can see from the figure, both Caligula and Nero share many alleles IBD. As an example of summarising a pairwise relationship, Germanicus and Agrippina Maior, parents of Caligula and of Nero's mother, have probabilities of sharing none and one of their genes IBD equal to  $\kappa_0=0.9375 $ and  $\kappa_1= 0.0625$.  The IBD pattern distribution, and coefficients like these extracted from it, very compactly yet exactly capture the very complex multi-generational story told by the pedigree.

\section{Computations for DNA mixtures}
\label{sec:specific}

\subsection{Likelihoods and Bayesian networks}

Under our universal assumption that we are using unlinked autosomal markers, and all individuals are drawn from a population in Hardy-Weinberg equilibrium, it is clear that genotypes $\n$ and peak heights $\z$ are independent across markers. Then the distributions $p(\n)$ and $p(\z|\n)$ are products over markers -- and therefore so are likelihood ratios computed from them. We therefore consider each marker separately, and assume a single marker for the rest of this section. The likelihood for observed peak heights $\z$ is of course
\begin{equation}\label{eq:pz}
p(\z) = \sum_{\n} p(\n)p(\z|\n),
\end{equation}
regarded as a function of the parameters in the two model component distributions. 

For the peak height model introduced and investigated by \textcite{cowell:etal:15}, $p(\z|\n)$ is defined as follows. As in \textcite{cowell:etal:15} it is convenient both algebraically and computationally to represent genotypes $\n$ by allele count arrays: $n_{ia}=0,1,2$ being the number of allele $a$ in the genotype for individual $i$. 

The peak heights $Z_{a}$  at allele $a$ are  gamma distributed, conditional on $\n$,
\begin{equation}\label{eq:gammadist}
 p(Z_{a}=z_a|\n) = g(z_{a};\rho D_a(\phi,\xi,\n),\eta),
\end{equation}
independently for each $a$, where $g$ denotes the gamma density function, $\rho$ quantifies the total amount
of DNA in the mixture before amplification, $\xi$ denotes the mean stutter proportion, $\eta$ the scale (not rate), $D_a(\phi,\xi, \n)= (1-\xi)\sum_{i}\phi_in_{ia} + \xi\sum_{i}\phi_in_{i,a+1}$ the effective allele counts taking into account the stutter artefact and  $\mathbf{\phi}=(\phi_i)_{i=1}^I$ denotes the proportion of DNA from individual $i$ in  the mixture.

For given genotypes $\n$, given parameters  $\psi=(\rho, \xi, \eta, \phi)$, the conditional distribution  for the  peak heights $\z=(z_{a})_{a=1}^A$ factorizes over alleles $a$

\begin{equation}
p(\z|\n) = p(\z|\n;\psi)
=\prod_a L_{a}(z_{a})
\end{equation}
where

\begin{align*}
L_{a} (z_{a})= \left\{ \begin{array}{cr} g(z_{a};\rho D_a(\phi,\xi,\n),\eta)& \mbox{ if $z_{a}>C$}\\G(C;\rho D_a(\phi,\xi,\n),\eta)&\mbox{otherwise,}\end{array}\right.
\end{align*}
with $G$ denoting the gamma distribution function and $C$ the detection threshold.

For a given hypothesis $\H$ about the joint distribution $p(\n)$ of the genotypes of the $I$ individuals, the full likelihood is obtained by summing over all possible combinations of genotypes
to give $p(\z)=p(\z;\psi)$ as in (\ref{eq:pz}).

Unless both the number of mixture contributors and the number of alleles are very small, computing the sum in (\ref{eq:pz}) is a formidable task, often an intractable one. The key observation in \textcite{graversen:lauritzen:comp:13}, exploited in the {\tt DNAmixtures} package, is that if genotypes $\n$ are encoded as allele count arrays $(n_{ia})$ giving the number of alleles of type $a$ for individual $i$, and the joint distribution $p(\n)$ factorised into conditional distributions sequentially over $a$ for each $i$, then $p(\n)$ has the structure of a Bayesian network (BN) with considerable sparsity.

Computation of $\sum_{\n} p(\n)p(\z|\n)$, which is the expectation over the BN distribution for $\n$ of the function $p(\z|\n)$, is then exactly the kind of task performed by a BN probability propagation algorithm \cite{laur/spieg}.
We follow the {\tt DNAmixtures} formulation, including the allele count array representation of genotypes; for more computational detail see \textcite{graversen:lauritzen:comp:13}. 

The IBD pattern distribution formulation helps to create methodology that considerably extends that described in \textcite{green:mortera:17}. That paper introduced four approaches to adapting the Bayesian network computation in {\tt DNAmixtures} to deal with paternity testing (where the putative father is a contributor to the mixture, and with or without the mother's genotype profile being available in addition to the child's). Three of these: ALN (additional likelihood node), RPT (replace probability tables) and MBN (meiosis Bayesian network) involve modifying the Bayesian network and are fast and essentially exact, the other, WLR (weighted likelihood ratio) is slow and approximate but easy to code. 

It turns out that the RPT method is easiest to adapt to the case of relationships that are more complex or involve more contributors. This entails replacing the default genotype conditional probability tables (CPTs), representing independent multinomial draws from the gene pool, by tables that encode assumed relationships and condition on any observed genotypes.
All genotypes are determined by the values of the draws from the gene pool and the meiosis pattern, so by including the gene pool draws and the meiosis pattern as nodes in the BN, the CPTs for the individual genotype arrays consist only of 0's and 1's; they are then easy to construct, and can be stored compactly and processed efficiently in some BN engines, including {\tt Hugin}.

Rather than define this process formally in the cumbersome and unilluminating algebra needed for generality, here we work through an example in detail. We begin by giving a constructive definition of the joint distribution of a family's genotypes, using the IBD pattern distribution describing the family; this is useful in its own right, but it could also be used explicitly in simulating cases for testing purposes, and it also helps to motivate how we can construct CPTs for allele count arrays.

\subsection{The joint distribution of genotype profiles when contributors are related}
\label{sec:simrelprof}

Here we show how the IBD pattern distribution, together with the allele frequencies for each marker, determines the joint distribution of genotype profiles. For expository purposes only, we will give a constructive specification, as if it was intended to simulate profiles from the correct distribution. STR markers are at unlinked loci, so profiles at different markers are independent.

Consider first the 4-individual family in Table \ref{IBDpatt}(c), which shows 2 IBD patterns, with probabilities 1/2 each. For each marker, independently, we proceed in three stages:
\begin{enumerate}
\item Choose a pattern (row of the table) at random, with the indicated probability.
\item Draw alleles from the gene pool, according to the allele frequencies, independently for each distinct label in the pattern -- in this case there are 5 distinct labels, 1,2,\ldots,5 in each pattern, so we draw alleles $a_1,a_2,\ldots,a_5$.
\item Assign gene pool draws to form genotypes for the individuals using the labels -- so here we set $\text{Fgt}=(a_1,a_2),\text{Mgt}=(a_3,a_4), \text{Cgt}=(a_1,a_3)$ and $\text{GFgt}=$ either $(a_1,a_5)$ or $(a_2,a_5)$ for rows 1 and 2 respectively.
\end{enumerate}

The general formulation should be clear from this example. Given the IBD pattern distribution, an allele will be drawn from the gene pool for each distinct label among the patterns, using the population allele frequencies for the marker in question. A pattern is drawn from the IBD pattern distribution, and the genes and hence genotypes for all individuals obtained by selecting the corresponding gene pool draws. The IBD pattern distribution applies equally for all markers, but the drawing of the alleles from the gene pool and of the IBD pattern will be independent for each marker. It is important to note that this procedure generates the genotypes for all the individuals of interest \emph{simultaneously}, rather than sequentially from parents to child.

Pseudo-codes for this algorithm, viewed as a method for simulating genotypes for members of a family, and those of the next two sections are given in Supplementary information, Section 1, and all are implemented in open-source code in the \KinMix\ package.

\subsection{Computing CPTs for related contributors}
\label{sec:cptrelated}

The case where no individuals are genotyped, so we are simply modelling family relationships among some or all of the mixture contributors is very straightforward. As in the previous section, we use the IBD pattern distribution directly. Continuing the Table \ref{IBDpatt}(c) example, with the 4 individuals labelled $i=1,2,3,4$, the nodes of the required BN represent $\{n_{ia}, i=1,2,3,4; a=1,2,\ldots,A\}$, $\{m_{ja}, j=1,2,3,4,5; a=1,2,\ldots,A\}$, $\{T_{ja}, j=1,2,3,4,5; a=2,\ldots,A-1\}$, and $s$. Here $n_{ia}=0,1,2$ is the number of alleles $a$ for individual $i$, $m_{ja}=0,1$ the numbers of $a$ for the $j$th draw from the gene pool, and the  $T$ are cumulative sums of the $m$ (cumulative over $a$). Again, pseudo-code for this algorithm is given in Section 1 of the Supplementary information, following the same pattern as that of the previous section, but at each node, instead of generating an allele count, we compute its conditional distribution given its parents in the graph in the form of a table.

\begin{figure}[htbp]
\centering
\resizebox{15cm}{!}{\includegraphics{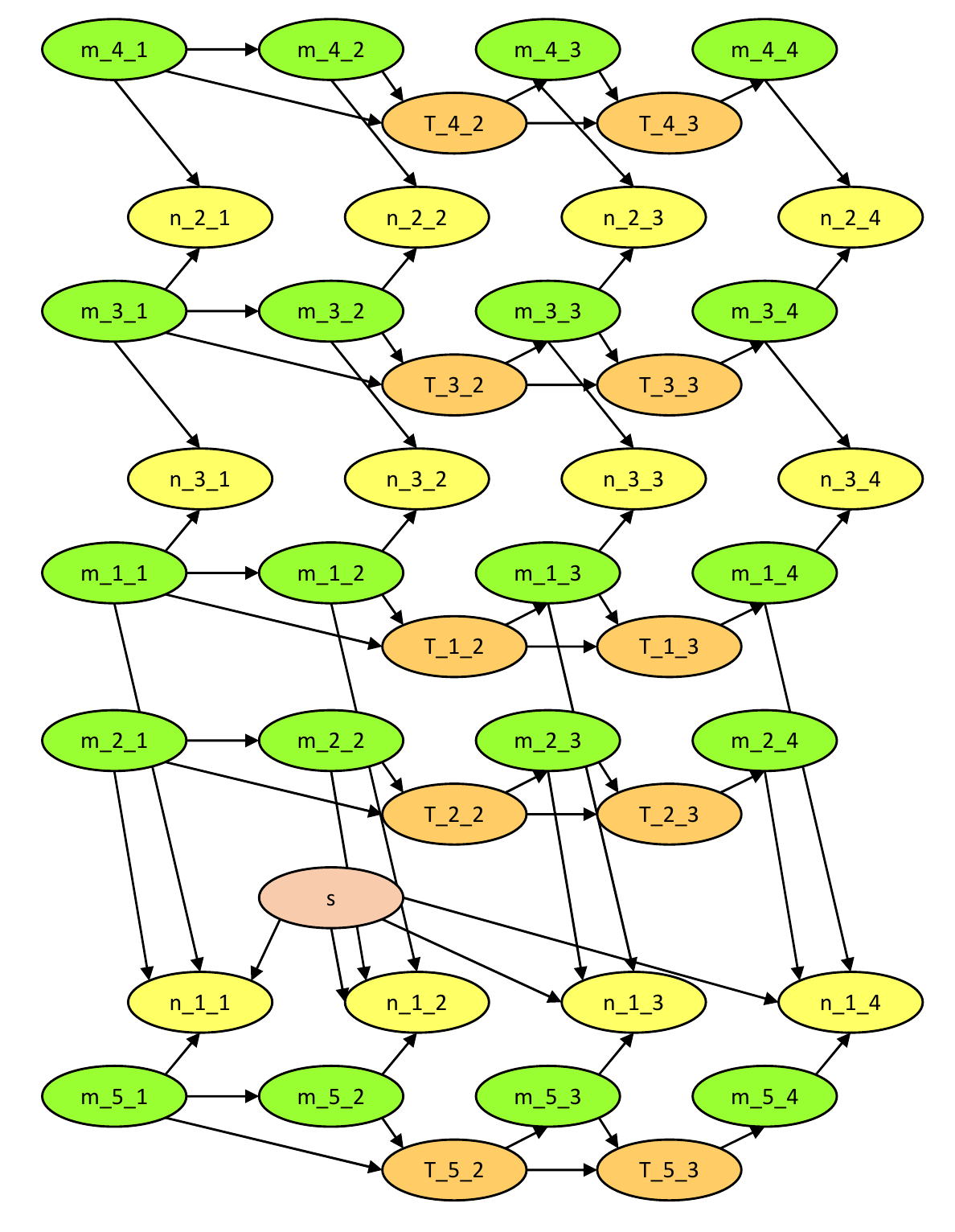}}
\caption{The typical structure of a BN for the joint distribution of several genotypes using the IBD pattern distribution: 3 genotypes (those for the Grandfather, Mother and Child in Table \ref{IBDpatt}(c)), derived from 5 draws from the gene pool, with 4 alleles. The node {\tt s} represents the IBD pattern.\label{fig:BN}}
\end{figure}

The typical structure of the BN representing this model when the gene pool draws and individual's genotypes are represented by allele count arrays is shown for a small example in Figure \ref{fig:BN}; this corresponds to the situation in Table \ref{IBDpatt}(c), but including just 3 of the 4 individuals in this family, the Grandfather, Mother and Child, here numbered 1, 2, 3 respectively, and assuming just 4 allelic values. Note that there are directed edges from $m_{ja}$ to $n_{ia}$ if and only if the pattern label $j$ appears among the IBD patterns for individual $i$. 
In general the graph depends on the number of individuals in the family, the number of alleles, and the IBD pattern distribution.

\subsection{Conditioning on typed relatives}
\label{sec:cptconditioning}

The case where some individuals in the family have been typed is a little more complicated; again similar reasoning applies to both simulation of genotypes, and computations of their CPTs.

Consider the 4-individual family again, and suppose that the Father and Mother are mixture contributors, and the Child and Grandfather are typed, with genotypes ($a$,$b$) and ($b$,$c$) respectively, where $a$, $b$ and $c$ are distinct alleles. 

\begin{table}[ht]
\caption{The worked example.\label{worked}}
\begin{center}
\begin{tabular}{cccccccc|c|cc}
\hline
\multicolumn{2}{c}{Fgt} & \multicolumn{2}{c}{Mgt} & \multicolumn{2}{c}{Cgt} & \multicolumn{2}{c|}{GFgt} & p(Cgt,GFgt) & Fgt & Mgt \\
\hline
 1 & $\ub{2}$ & 3 & $\ub{4}$ & 1 & 3 & 1 & 5 & 0 & & \\
 1 & $\ub{2}$ & 3 & $\ub{4}$ & 3 & 1 & 1 & 5 & $0.125q_aq_bq_c$ & ($b$,?) & ($a$,?) \\
 1 & $\ub{2}$ & 3 & $\ub{4}$ & 1 & 3 & 5 & 1 & 0 & & \\
 1 & $\ub{2}$ & 3 & $\ub{4}$ & 3 & 1 & 5 & 1 & 0 & & \\
 1 & 2 & 3 & $\ub{4}$ & 1 & 3 & 2 & 5 & $0.125q_aq_b^2q_c$ & ($a$,$b$) & ($b$,?) \\
 1 & 2 & 3 & $\ub{4}$ & 3 & 1 & 2 & 5 & $0.125q_aq_b^2q_c$ & ($b$,$b$) & ($a$,?) \\
 1 & 2 & 3 & $\ub{4}$ & 1 & 3 & 5 & 2 & $0.125q_aq_b^2q_c$ & ($a$,$c$) & ($b$,?) \\
 1 & 2 & 3 & $\ub{4}$ & 3 & 1 & 5 & 2 & $0.125q_aq_b^2q_c$ & ($b$,$c$) & ($a$,?) \\
\hline
& & & & $a$ & $b$ & $b$ & $c$ & $a\neq b \neq c\neq a$ & &\\
\hline
\end{tabular}
\end{center}
\end{table}

Our construction is shown schematically in Table \ref{worked}. Recalling that genotypes are unordered pairs of genes, we first expand the IBD pattern distribution table by explicitly laying out all possible permutations of the allele labels for the two \emph{typed} individuals, giving equal probability to each; this is to expedite checking for matches for each label. Matching these four allele labels onto their observed values $a$,$b$,$b$,$c$ respectively allows us to map allele labels onto actual allele values, for each of the permuted patterns. It also reveals that some permuted patterns cannot generate the observed alleles. For example the first row is impossible, given the observed genotypes, since the allele label 1 cannot simultaneously map onto the distinct alleles $a$ and $b$, while the second row is possible, with the mapping $1\mapsto b$, $3\mapsto a$, $5\mapsto c$. These mappings already determine some of the alleles of the Father and Mother, the mixture contributors. Those that are still not fixed are distinguished by underlining in bold in Table \ref{worked}. A column, headed $p(Cgt,GFgt)$, has been added to the table giving for each permuted pattern the probability of the typed individuals having the observed values. The final columns for the table summarise what is now known about the Father's and Mothers' genotypes for each permuted pattern, with ? denoting a draw from the gene pool. (In this example, all such draws are independent, since there is no duplication in allele labels for the Father and Mother). 

We now have all the information needed either to simulate Father and Mother genotypes, or to construct a BN for the genotype allele count arrays of the Father and Mother, in each case conditional on the genotypes of the Child and Grandfather. The BN `parents' for each count node consist of one node indexing the permuted pattern, together with nodes indicating the values of the draws from the gene pool required. The probability distribution over the permuted pattern node is modified from the `prior' (uniform) distribution by being conditioned on the typed genotypes, that is, it consists of the values of p(Cgt,GFgt) (see Table \ref{worked} in the case of our worked example), renormalised to sum to 1. 

\begin{table}[ht]
\caption{A simpler example: simple paternity testing.\label{paternity}}
\begin{center}
\begin{tabular}{cccccc|c|c}
\hline
\multicolumn{2}{c}{Fgt} & \multicolumn{2}{c}{Mgt} & \multicolumn{2}{c}{Cgt} & p(Mgt,Cgt) & Fgt \\
\hline
 1 & $\ub{2}$ & 3 & $4$ & 1 & 3 & 0 &  \\
 1 & $\ub{2}$ & 4 & $3$ & 1 & 3 & 0 &  \\
 1 & $\ub{2}$ & 3 & $4$ & 3 & 1 & 0 &  \\
 1 & $\ub{2}$ & 4 & $3$ & 3 & 1 & $0.25q_aq_bq_c$ & ($c$,?) \\
\hline
& &  $a$ & $b$ & $b$ & $c$ & $a\neq b \neq c\neq a$ & \\
\hline
\end{tabular}
\end{center}
\end{table}

It might be useful to point out that the approach we take to computing conditional genotype probabilities (as a crucial step on the way to delivering likelihood ratios) avoids any manual algebra, which is straightforward in simple cases but can be tedious and error-prone otherwise. Of course, it obtains the same answer. To see this, consider the familiar example of paternity testing given both mother's and child's genotypes. For the case where these two genotypes are $(a,b)$ and $(b,c)$ respectively, where again $a,b,c$ are all different, see Table \ref{paternity}, which is in exactly the same format as Table \ref{worked}. Simple algebra tells us that the Father must donate the $c$ allele to his child, and that his other allele is drawn from the gene pool, and this is exactly the answer that Table \ref{paternity} provides.

\section{Coancestry and uncertain allele frequencies}
\label{sec:mixrel}

We have so far focussed on the role in modelling DNA mixtures of close relationships, specified through family structures. In this section, we will briefly touch on the different situation of coancestry, or what we could call \emph{ambient relatedness}, where purportedly unrelated actors in fact have dependent genotypes because the population from which they are drawn exhibits high relatedness, for example through inbreeding. Just as with close relationships, these dependencies are driven by identity by descent, but the impact is somewhat different, because it applies generally to the whole population, and the dependence is usually substantially lower in magnitude.

In model-based inference from DNA profiles of STR markers, it has become routine to apply the `$\theta$ correction' of \textcite{djb/ran:fsi}. The scalar parameter $\theta$, the kinship coefficient or coefficient of ancestry, can be identified with Wright's measure of interpopulation variation $F_{ST}$ \cite{wright40,wright51}, and informally interpreted as the `proportion of alleles that share a common ancestor in the same subpopulation'. As discussed by \textcite{djb/ran:fsi}, the parameter $\theta$ arises in various models for dependent populations, for example the `island model' of partially-separated sub-populations.
That $\theta$ does not determine coefficients of identity was noted in Section \ref{sec:coefIBD}.

\textcite{green:mortera:09} observed that exactly the same probabilistic model for the joint distribution of multiple genes arises in a simple model for \textit{uncertainty in allele frequencies} (UAF), in which the true allele frequencies are treated as unknowns with a Dirichlet distribution and the database used for calculation regarded as a multinomial sample from these true frequencies. The parameter $\alpha=(1-\theta)/\theta$ is then the effective size of the database.
\textcite{green:mortera:09} also observed that this model is amenable to implementation as a BN, as an alternative to algebraic specification; this BN actually encodes a P\'olya urn scheme.

\subsection{Ambient relatedness for allele count arrays} 
\label{sec:ambient}

In discussion of \textcite{cowell:etal:15}, both Tvedebrink (modelling kinship) and Green (modelling uncertainty in allele frequencies) observe that when genotypes are represented by allele counts arrays $n_{ia}$, the number of alleles $a$ of individual $i$, this P\'olya urn scheme can be expressed through the recursion

\begin{align}
n_{1.} \sim & \text{DM}(2,(\alpha_a)_{a=1}^A) \nonumber\\
n_{i.} | (n_{j.})_{j=1}^{i-1} \sim & \text{DM}(2,(\alpha_a+n_{<i,a})_{a=1}^A) \label{eq:DM}
\end{align}
where $n_{<i,a} = \sum_{j=1}^{i-1} n_{ja}$ (etc.), and 
$\text{DM}$ denotes the Dirichlet--Multinomial distribution. See \textcite{green:cglmdiscn,tvedebrink}.
The Dirichlet--multinomial distribution is the straightforward generalisation of the Beta--binomial to more than 2 categories.
It is a Dirichlet mixture of multinomial distributions. $X\sim \text{DM}(n,(\alpha_a)_{a=1}^A)$ means
$$
P(X=x) 
=
\left\{\frac{n!}{\prod_a x_a!}\right\}\times 
\left\{\prod_a \frac{\Gamma(\alpha_a+x_a)}{\Gamma(\alpha_a)}\right\}
\times \frac{\Gamma(\sum_a \alpha_a)}
{\Gamma(\sum_a \alpha_a+n)},
$$
so long as $\sum_a x_a=n$.

Factorising the conditional distributions in (\ref{eq:DM}) over alleles, we find that individual allele counts have Beta--Binomial conditional distributions:
$$
n_{ia} | (n_{j.})_{j=1}^{i-1},\{n_{ib},b<a\} \sim \text{BB}((2-n_{i,<a}),(\alpha_a+n_{<i,a}),(\alpha_{>a}+n_{<i,>a})).
$$
%where $\alpha_{>a}=\sum_{b>a} \alpha_b$, $n_{i,\leq a}=\sum_{b=1}^a n_{ib}$, $n_{<i,>a}=\sum_{j<i}\sum_{b>a} n_{jb}$. 
The Beta--binomial distribution: $X\sim \text{BB}(n,\alpha,\beta)$ means
$$
p(X=x) 
={n \choose x} \frac{\Gamma(\alpha+x)\Gamma(\beta+n-x)\Gamma(\alpha+\beta)}
{\Gamma(\alpha)\Gamma(\beta)\Gamma(\alpha+\beta+n)}
$$
Note that $\text{BB}(1,\alpha,\beta)$ is just Bernoulli$(\alpha/(\alpha+\beta))$.

The {\tt KinMix} package includes functions for modifying genotype CPTs to model UAF and ambient IBD.

\textcite{tvedebrink:15}  give a fuller analysis, including a quantitative study of the implications. They show that relatedness/uncertainty in allele frequencies can \textit{increase or decrease} \LR's in identification tasks. Section 3 of the Supplementary information provides a list of software that handles DNA mixtures in the presence of coancestry.

\subsection{Coancestry and relationships together}

Particular casework may involve both close relationships and ambient relatedness (or uncertainty in allele frequencies). Modelling such a situation combines elements from Sections \ref{sec:cptrelated} and \ref{sec:ambient}. Full details are omitted, but the key algebraic manipulations are given in Supplementary information, Section 2. The algorithmic implications are that Binomial distributions in algorithms would be replaced by Beta-Binomials, with the meiosis pattern as an additional parent at each instance of the P\'olya urn. See also \textcite{cowell2016combining} for a different analysis of this combined model. 

At present, simultaneous modelling of close relationships and coancestry/uncertainty in allele frequencies is not implemented in the {\tt KinMix} package. Previous work with allele-presence data only \cite{green:mortera:09} and the work of \textcite{tvedebrink:15} show that the numerical difference in log likelihood ratios due to coancestry is rarely important in scientific terms.

\section{Setting parameters}
The methodology of this article is based on the joint probability model $p(\n,\z)=p(\n)p(\z|\n)$ for the genotype profiles and peak heights, and this distribution has a number of parameters, notably the population allele frequencies, the relative proportions of the contributions of the different contributors to the mixture, and the parameters describing the PCR process and the artefacts of measurement embedded in the \textcite{cowell:etal:15} peak height model. We do not prescribe any particular approach to setting these parameters when evaluating the likelihood, as this choice must be strongly influenced by regulation and practice in the judicial regime in which the analysis of the mixture is to be used, and the particular question that is being addressed.

Although Bayesian networks are a key concept in the computations we use, this does not mean that a Bayesian formalism is intended. In fact the \textcite{cowell:etal:15} model is presented as entirely frequentist, with the BN computations simply a device to compute an otherwise intractable likelihood function. The \texttt{DNAmixtures} package includes a function for maximising the likelihood as a function of the model parameters (the relative proportions of the contributions of the different contributors to the mixture, and the parameters describing the PCR process and the artefacts of measurement), and that function applies equally to a model modified using \texttt{KinMix}.

In principle a Bayesian analysis could be conducted, and the fast calculations of the likelihood that the packages provide would be an asset in implementing a Markov chain Monte Carlo (MCMC) sampler for posterior simulation, but we have not attempted this.  

Considering now only maximum likelihood estimation (MLE), when we are comparing alternative models for mixtures, the question arises under which model(s) should the likelihood be maximised? The answer depends on context and perspective. In a likelihood ratio test of $\Hp$ against $\H_0$, the respective likelihoods would each be maximised separately, and the ratio of the maximised values used as the test statistic. However, here the $\log_{10} \LR$ measures the weight of the evidence, and we are not appealing to statistical testing theory. In a criminal trial, depending on jurisdiction, custom might suggest or dictate choosing parameter values for both numerator and denominator that minimise the ratio, or perhaps maximise the denominator, in line with the presumption of innocence of the defendant (\emph{in dubio pro reo}). In a civil case, say a paternity suit, the notion of defendant hardly applies.

Our experience has been that in many contexts, choice between these approaches makes little difference to the values of likelihood ratios, or their interpretations; we give a numerical example of this below. However, this cannot be a general conclusion; we anticipate, for example, that in comparing alternative hypotheses about the relatedness of mixture contributors to each other, parameter choice could have more impact.

Our numerical illustration revisits the Italian singer case, used for motivation in \textcite{green:mortera:17} (sections 4 and 5). This is a case of paternity testing, where the putative father is represented in the evidence only as a contributor to a mixture, assumed to be of 2 contributors, denoted $U_1$ and $U_2$; this is because the claim of paternity only arose some years after the putative father's death and the remains available had been corrupted. The child's genotype Cgt is available, and we presented weights of evidence for paternity, with and
without using also the mother's genotype Mgt. The hypotheses are $\Hp$, that the major contributor to the mixture is the father, against $\H_0$, that neither contributor to the mixture is related to the child (or mother).

Table \ref{tab:param} illustrates that under either $\H_0$ or $\Hp$ the MLEs ($\widehat{\psi}_0$, $\widehat{\psi}_p$, respectively) of the parameters $\psi=(\rho, \eta, \xi, \phi_{U1}, \phi_{U2})$  do not vary substantially, as mentioned above.
Table \ref{tab:LRparam} gives the likelihood ratios in favour of paternity without and with information  on Cgt's mother's genotype Mgt: a) when using the MLEs computed under $\H_0$ in both numerator and denominator of the LR (columns 2 and 3) and b)  when using the MLEs computed separately  under $\H_p$  and $\H_0$ (columns 4 and 5). Adopting a) or b) makes only immaterial differences to the LRs (less than 2\%), in the context where the LR in favour of paternity is already more than 250,000, and including also the information about the mother's genotype increases this at least 500-fold.  

\begin{table}[htb]
  \centering
   \caption{Maximum likelihood estimates in the Italian singer paternity case; the parameters are those in the \textcite{cowell:etal:15} peak height model.}
  \label{tab:param}
    \vspace{2mm}
  \begin{tabular}{|l|c|c|c|c|c|}
\hline
MLEs & $\rho$ & $\eta$ & $\xi$ & $\phi_{U1}$ & $\phi_{U2}$ \\
   \hline
 $\H_0$ -- baseline &  0.718 &  1124  &  0.006643 &  0.9783 &  0.02166 \\
 $\Hp$ with Cgt known &  0.745&  1083&  0.006527 &  0.9797 &  0.02034 \\
 $\Hp$ with Cgt \& Mgt known &  0.745 &  1083 &  0.006526 &  0.9797 &  0.02035 \\
  \hline
  \end{tabular}
    \end{table}
			
	\begin{table}[htb]
  \centering
   \caption{Likelihood ratios and their logarithms, in the Italian singer paternity case}
  \label{tab:LRparam}
	\vspace{2mm}
  \begin{tabular}{|l|c|c|c|c|}
\hline
Likelihood ratios & \multicolumn{2}{c|}{$\psi_p=\psi_0=\widehat{\psi}_0(z)$} & \multicolumn{2}{c|}{$\psi_p=\widehat{\psi}_p, \psi_0=\widehat{\psi}_0$} \\
\hline
$\Hp$ vs. $\H_0$ & $\LR$ & $\log_{10}(\LR)$ & $\LR$ & $\log_{10}(\LR)$  \\
   \hline
 with Cgt  &  $266100$ &  $5.425$ &  $270100$ &  $5.432$  \\
 with Cgt \& Mgt known &  $143.5\times 10^6$ &  $8.157$ &  $145.7\times 10^6$ &  $8.163$  \\
  \hline
  \end{tabular}
\end{table}

\section{Software: the KinMix package}

The modelling and methods introduced in this paper form the basis for an \textbf{R} package \texttt{KinMix} \cite{kinmix}, that extends the package \texttt{DNAmixtures} \cite{graversen:package:13}. As with \texttt{DNAmixtures}, therefore, the package relies on the \texttt{Hugin} system for probabilistic expert systems, accessed via the \texttt{Rhugin} package. The model for peak heights given genotypes $p(\z|\n)$, together with the treatment of artefacts such as drop-out and stutter, are exactly as in \textcite{cowell:etal:15}. The \texttt{KinMix} package provides functions for constructing IBD pattern distributions from pedigree information, and using these pattern distributions to modify the default \texttt{DNAmixtures} genotype profile distribution $p(\n)$ (in which untyped individuals are assumed unrelated draws from a specified gene pool), to allow for related contributors.

\texttt{KinMix} also provides facilities from simulating genotype profiles from groups of individuals with specified joint relationships, making graphical displays of joint relationships, and many other utilities. In addition to the package manual pages in standard \textbf{R} format, a tutorial user guide is available as \textcite{kinmix-userguide}.

\texttt{KinMix} inherits from \texttt{DNAmixtures} the representation of genotypes via allele count arrays, which is important in saving both computation time and computer memory, and allows the modelling of mixtures with as many as 5 or 6 untyped individuals, depending on the details of the case, especially the numbers of alleles, and on the memory available. Although building in relationships between individuals does increase both time and space requirements, examples in this paper demonstrate that quite complex problems can be considered. Under standard assumptions, once parameters are fixed, likelihoods and likelihood ratios in our mixture models factorise over markers. Further, in the case of related contributors, logarithms of likelihoods and likelihood ratios are weighted averages of those obtained from the individual IBD patterns. The weights are the probabilities of the patterns, the posterior probabilities in the case where some of the relatives have been typed. An option in \texttt{KinMix} allows exploiting these facts, with the effect of considerably reducing the storage requirements when many markers are involved, since separate BNs are used for each marker/IBD pattern combination.

%\par\noindent\textbf{Timings\\
%Max size of problem, with examples}

\section{Simulations}
\label{sec:simulations}

In this section, we examine the performance of $\log_{10}$-likelihoods, based on the \textcite{cowell:etal:15} model, at discriminating between different joint relationships, in a study using simulated electropherogram (EPG) and genotype data. Each simulated data set consists of genotype profiles generated from a prescribed `True' model, using the generative model in Section \ref{sec:simrelprof}; from each we generate artificial EPG data using the \texttt{pcrsim} \cite{pcrsim} package, which simulates the DNA amplification process. 
These PCR simulations were based on using the AmpF$\ell$ STR\texttrademark\ SGM Plus\texttrademark\ PCR Amplification Kit, with the Norwegian SGM database, on 10 STR markers, together with 
Amelogenin, all available in \texttt{pcrsim}. The numbers of cells amplified varied from 200 to 25\footnote{Other PCR parameters were as defined by the \texttt{pcrsim} command \texttt{simPCR(data=res, pcr.prob=1, pcr.cyc=28, vol.aliq=20,   vol.pcr=50, sd.vol.pcr=1)}.},

The resulting data are analysed using \texttt{KinMix} \cite{kinmix}, under a variety of assumed models. In each case, we compute the log likelihood ratio against a baseline model that assumes the same number of unrelated contributors. In all examples, the focus of interest is on the $\log_{10} \LR$ from the peak height data \emph{given} the stated available genotypes, that is, on the additional information about the question of interest attributable to the mixture data. 
These ratios are tabulated or graphed for a small number of replicates\footnote{These are independent identically distributed repeats of the PCR simulation, not replicates in the sense of DNA replication.} of the PCR process, for each of a small number of independently generated genotype profiles, thus giving an idea of the variation attributable to these two sources. Parameters for the \texttt{DNAmixtures} peak height model are estimated by maximum likelihood using that package, assuming the appropriate number of unrelated contributors, estimated separately on each simulated EPG.

Each analysis in the following experiments involves specifying the true joint relationship between the actors involved, the hypothesised relationship(s), which of the actors contribute to the mixture, and which if any of the actors are genotyped.

\subsection{Study 1: Two-way relationships}
\label{sec:ex1}

%\emph{These runs are not easily replicated - the run used here is saved in 202005071502.RData}

In this experiment, we study DNA mixtures with two contributors, and no other typed actors. We consider 5 possible familial relationships between the two contributors; for each relationship, we simulate EPG data and fit mixture models hypothesising each of the 5 relationships in turn. In Table \ref{tab:2way}, we report the median $\log_{10} \LR$ for each of the hypothesised models, over 4 replicated EPGs for each of 4 replicated genotypes. In each case the EPG data are simulated with 150 cells from the major contributor and 50 from the minor one.

 \begin{table}[htb]
 	\centering
 	\caption{Study 1: Median $\log_{10} \LR$ over 4 replicated EPGs for each of 4 replicated genotypes, for 5 hypothesised models.}
 	\label{tab:2way}
	\vspace{2mm}
 	\begin{tabular}{l|ccc|rrrrr}
	\hline
 		\multicolumn{1}{c|}{true}& & & & \multicolumn{5}{c}{hypothesised} \\
 		 & $\kappa_0$ & $\kappa_1$ & $\kappa_2$ & parent--child &      sibs  & half-sibs  &    cousins & half-cousins \\
 		\hline
% 		parent--child & 0 & 1 & 0 &     3.37 &       2.11 &       2.13 &       1.27 &       0.72 \\
% 		sibs & 0.25 & 0.5 & 0.25 &   $-$18.25 &       2.40 &       1.52 &       0.96 &       0.53 \\
% 		half-sibs & 0.5 & 0.5 & 0 &     $-\infty$ &        $-$0.32 &       0.46 &       0.43 &       0.28 \\
% 		cousins & 0.75 & 0.25 & 0 &       $-\infty$ &     $-$1.25 &       0.03 &       0.16 &       0.13 \\
% 		half-cousins & 0.875 & 0.125 & 0 &   $-\infty$ &     $-$1.58 &      $-$0.40 &      $-$0.05 &       0.01 \\
%	  parent--child & 0 & 1 & 0 &    2.956  &   2.232   &       1.822  &      1.040 &       0.570\\
% 		sibs & 0.25 & 0.5 & 0.25 &   $-\infty$ &       2.350      &    1.697    &    1.147 &       0.652 \\
% 		half-sibs & 0.5 & 0.5 & 0 &     $-\infty$ &       $-$0.914    &      0.172    &    0.245 &       0.170 \\
% 		cousins & 0.75 & 0.25 & 0 &       $-$35.512  &  $-$1.191    &      0.070   &     0.194 &       0.152 \\
% 		half-cousins & 0.875 & 0.125 & 0 &   $-\infty$ &   $-$2.045   &      $-$0.513   &    $-$0.097 &        $-$0.010 \\		\hline
 	  parent--child & 0 & 1 & 0 &  2.360 & 1.927 & 1.873 & 1.129 & 0.636 \\

 		sibs & 0.25 & 0.5 & 0.25 &    $-4.089$ & 1.812 & 1.294 & 0.748 & 0.409 \\
 		half-sibs & 0.5 & 0.5 & 0 &   $-22.508$ & $-0.831$ & 0.474 & 0.355 & 0.198 \\
 		cousins & 0.75 & 0.25 & 0 &   $-\infty$ & $-0.472$ & 0.219 & 0.320 & 0.213 \\
 		half-cousins & 0.875 & 0.125 & 0 &  $-\infty$ & $-1.939$ & $-0.607$ & $-0.097$ & 0.004 \\		\hline
	\end{tabular}  
 \end{table}
 
\begin{figure}[htbp]
 	\begin{center}
 		\includegraphics{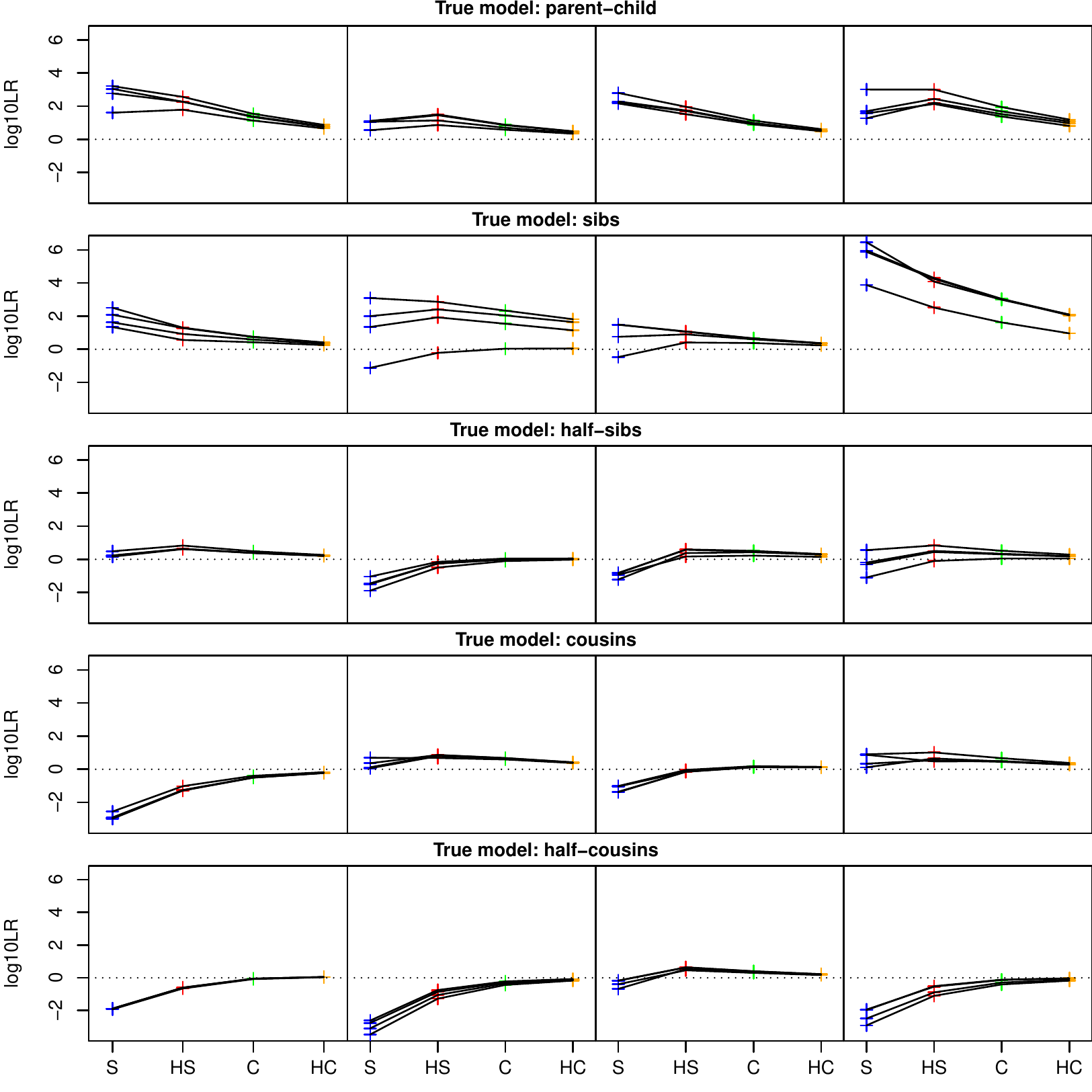}
		\caption{Study 1: Variation in the $\log_{10} \LR$s across replicate genotype profiles and EPGs for the two-way example.
The rows of the figure correspond to the true relationships, and the columns to the 4 replicate genotype profiles. Within each panel $\log_{10} \LR$ is plotted against hypothesised relationship (from left to right: sibs (S), half-sibs (HS), cousins (C) and  half-cousins (HC), respectively). Lines join values corresponding to the same EPG replicate.
The $\log_{10} \LR$s when parent--child is hypothesised are suppressed from the Figure as they take extreme values, as can be seen in the parent--child column of Table \ref{tab:2way}.}
 		\label{fig:2way}
 	\end{center}
 \end{figure}

The variation in these $\log_{10} \LR$s across replicate genotype profiles and EPGs is depicted in Figure \ref{fig:2way}.  The rows of the figure correspond to the true relationships, and the columns to the 4 replicate genotype profiles. Within each panel we see a colour-coded diagram showing the variation in $\log_{10} \LR$ over the 4 EPG replicates. 
The $\log_{10} \LR$s when parent--child is hypothesised are suppressed from the Figure as they take extreme values, as can be seen in the parent--child column of Table \ref{tab:2way}.

The highest median $\log_{10} \LR$s are all found down the diagonal, so that if we select a model on the basis of the largest, then on average, we correctly identify the true model in all 5 cases. This is most pronounced when the true model is parent--child, a pattern that is perhaps to be expected given the $\kappa$ coefficients, also tabulated in Table \ref{tab:2way}. Some of the other relationships are harder to distinguish. Also, recall that many relationships, like half-sibs, aunt/uncle, grandparent \etc\ have identical IBD pattern distributions and $\kappa$ coefficients. Perhaps a more meaningful interpretation is that on an individual-replicate basis, for the 5 true relationships, 
%in 8, 13, 7, 5 and 10 
%in 12, 14, 3, 7 and 9 
in 11, 11, 9, 4 and 12 out of the $16=4\times 4$ replicates respectively, the correct model was identified. 
  
   When there is additional information, such as the genotypes of  individuals potentially related to mixture contributors the evidence becomes much stronger, as we will see in some of the following examples. 

\subsection{Study 2: Three-way relationships}

This experiment is exactly similar to the previous one, except now we consider relationships between 3 related contributors. The five considered relationships are respectively trio (mother, father and child); a mother and two children; 3 sibs; 3-cousins-cyclic, 3-cousins-star; the last two are defined and illustrated in Section \ref{sec:joint}. Again, there are  no typed actors.
In each case the EPG data are simulated with 100, 50 and 25 cells from the three contributors.   

\begin{table}[htb]
	\centering
	\caption{Study 2: Median $\log_{10} \LR$ over 4 replicated genotypes by 4 replicated EPGs   in  simulations  of a DNA mixture for each 3-way relationship.}
	\label{tab:3way}
	\vspace{2mm}
	\begin{tabular}{l|rrrrr}
	\hline
	\multicolumn{1}{c|}{true}	& \multicolumn{5}{c}{hypothesised}\\
	              &       trio & mother, 2 kids &     3 sibs & 3-cousins-cyclic & 3-cousins-star \\
\hline
%trio &       2.00 &      $-$7.30 &       1.16 &       1.63 &       1.11 \\
%mother, 2 kids  &      $-$9.83 &       4.32 &       2.51 &       2.51 &       1.93 \\
%3 sibs &      $-$0.69 &      $-$1.50 &       4.27 &       2.25 &       1.88 \\
%3cousins-cyclic &     $-$72.32 &     $-$57.25 &     $-$46.23 &       0.32 &       0.16 \\
%3cousins-star &     $-\infty$ &     $-\infty$ &    $-\infty$ & 3.9 $\times 10^{-4}$ &     $-$0.045 \\
%trio & 1.178 & $-9.647$ & 1.740 & 1.853 & 1.410 \\
%mother, 2 kids  & $-3.132$ & 4.631 & 3.615 & 2.541 & 2.373 \\
%3 sibs & $-2.866$ & $-1.118$ & 5.903 & 3.259 & 2.805 \\
%3cousins-cyclic & $-20.295$ & $-10.304$ & $-5.295$ & 0.985 & 0.967 \\
%3cousins-star & $-\infty$ & $-\infty$ & $-\infty$ & 0.090 & 0.281 \\
trio &   2.772 & $-15.903$ &  1.772 & 1.822 & 1.310 \\
mother, 2 kids  &  2.032 &  5.280 & 4.260 & 2.929 & 2.230 \\
3 sibs & $-$6.446 & $-$3.800 & 3.584 & 2.417 & 1.917 \\
3cousins-cyclic & $-$14.354 & $-$20.380 & $-$7.308 & 0.696 & 0.585 \\
3cousins-star & $-$20.741 & $-$20.325 & $-$5.740 & 0.502 & 0.540 \\
\hline
\end{tabular}  
\end{table}

Table \ref{tab:3way} gives the median $\log_{10} \LR$ relative to the baseline (three unrelated contributors) for comparing each combination of true and hypothesised relationship model, over 4 replicated genotypes by 4 replicated EPGs. 
The highest median $\log_{10} \LR$s are again all found down the diagonal, implying that each of the 5 relationships are correctly identified on average. This effect is strongest when the relationships are mother and 2 children, or 3 sibs.  In these two scenarios the 3 contributors all have a close pairwise relationship, whereas in the trio, the mother and father are unrelated. Also, the 3 cousins scenarios have more distant relationships among each other, and of course have identical pairwise relationships, so are  harder to distinguish. These factors explain the asymmetry in the table.
On a per-replicate basis, for the 5 true relationships, in respectively 
%9, 11, 11, 14 and 8 
%6, 9, 15, 8, 11 
12, 7, 12, 14, 11 out of the $16=4\times 4$ replicates, the correct model was identified.  

\begin{figure}[htbp]
	\begin{center}
		%\resizebox{18cm}{!}{
		\includegraphics{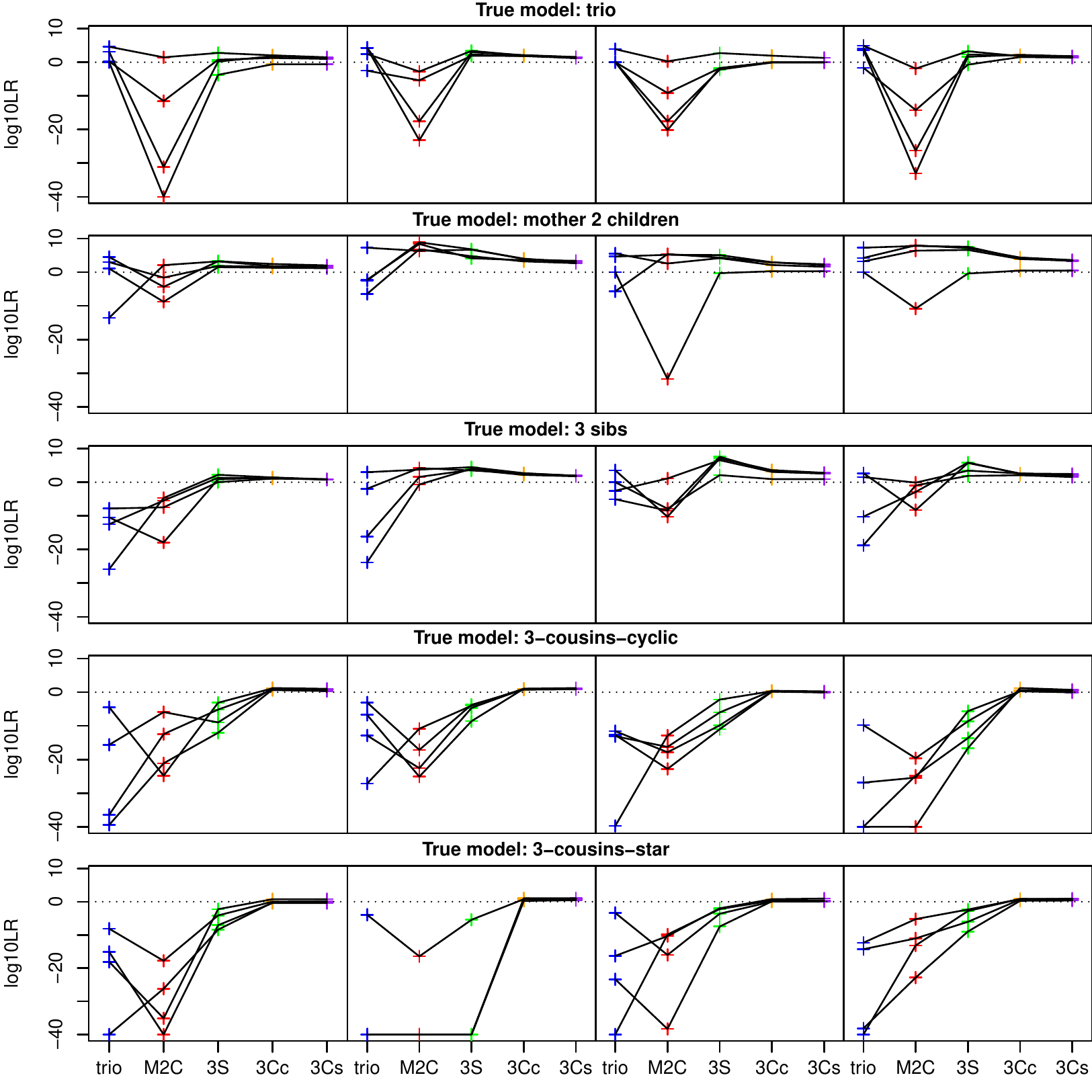}
		%}
		\caption{Study 2: Variation in the $\log_{10} \LR$s across replicate genotype profiles and EPGs for the three-way example.
The rows of the figure correspond to the true relationships, and the columns to the 4 replicate genotype profiles. Within each panel $\log_{10} \LR$ is plotted against hypothesised relationship (from left to right: trio, mother and two children (M2C), 3 sibs (3S), 3-cousins-cyclic (3Cc), 3-cousins-star (3Cs), respectively). Lines join values corresponding to the same EPG replicate. All values are truncated below at $-40$ before plotting.}
		\label{fig:3way}
	\end{center}
\end{figure}

Figure \ref{fig:3way} shows the variation in $\log_{10} \LR$s within the genotypes and across the 4 replicated  EPGs in each row  panel for 5 different 3-way relationships. As in the previous section, each row corresponds to a true relationship, and the four panels to the genotype profile replicates. 

%\todo{This information might be useful elsewhere: The time to do the entire analysis starting  from the simulations with all 16 replicates  and 5 scenarios  is 1.74 hours.}

\subsection{Study 3: Three-way relationships, with a relation genotyped}

Here we consider 4 brothers, and DNA mixtures simulated from a true model in which three of the brothers are contributors. We consider testing $\Hp$: 3 brothers contributed to the mixture \textit{vs.} $\H_0$: the 3 contributors are unrelated, and drawn from the gene pool. We perform this test with and without the assumption that the 4th brother is genotyped, yielding genotype Bgt, and as usual we generate 4 replicate joint genotype profiles, and 4 replicate EPGs for each. In each case the EPG data are simulated with 200, 100 and 50 cells from the three brothers contributing to the mixture; the same EPGs are used for the analyses without and with the 4th brother's genotype.

This kind of case can arise when brothers are engaged in a joint criminal activity and DNA might be found on, \eg\ a get-away car, balaklava,  banknote, crowbar, or gun. 

\begin{figure}[htbp]
	\begin{center}
		\resizebox{6.5in}{!}{\includegraphics{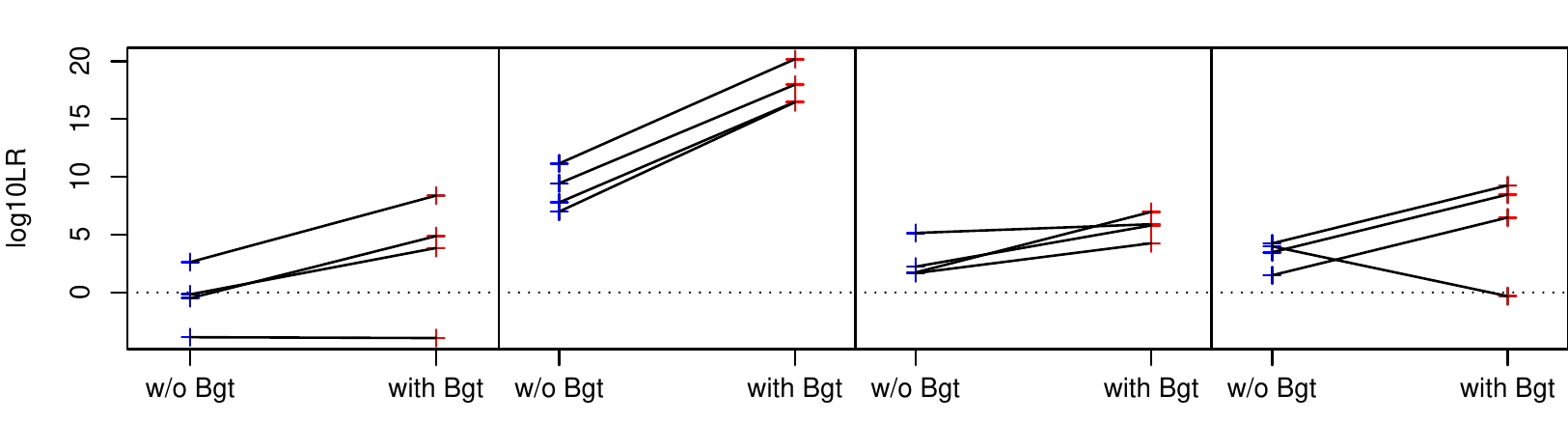}}
		\caption{Study 3: $\log_{10} \LR$ for $\Hp$ \textit{vs.} $\H_0$ in the case of 4 brothers, without and with the 4th brother genotyped. The 4 panels display results for the 4 replicate genotype profiles, the lines join results for the 4 replicate EPGs.}
		\label{fig:4sibs}
	\end{center}
\end{figure}

The $\log_{10} \LR$ results are shown by replicate in Table \ref{tab:4sibs} and Figure \ref{fig:4sibs}. Note that in most but not all of the $16=4\times 4$ replicates, there is much greater weight of evidence that the 3 brothers are in the mixture when the 4th brother's genotype is available.

\begin{table}[htb]
	\centering
	\caption{Study 3: The contributors to a mixture are 3 sibs. $\log_{10} \LR$ for testing whether the contributors are 3 sibs or not, i.e. $\Hp$: 3 sibs contributed to the mixture \textit{vs.}  $\H_0$:  the 3 contributors are unrelated. We also have the genotype Bgt of a 4th sib. The analysis is replicated $4 \times 4$ times.}
	\vspace{2mm}
	\label{tab:4sibs}
	\begin{tabular}{c|rrrr|rrrr}
	\hline
	& 	\multicolumn{4}{c|}{without Bgt} &	\multicolumn{4}{c}{with Bgt}\\
	\hline
EPG &	\multicolumn{4}{c|}{genotype profile} &	\multicolumn{4}{c}{genotype profile}\\
replicate & 1 & 2 & 3 & 4 & 1 & 2 & 3 & 4 \\
\hline
%1 & 6.26 & 7.30 & 2.87 & 5.78 & 14.16 & 13.32 & 6.01 & 14.04\\
%2 & 7.20 & 5.75 & 5.80 & 2.44 & 14.34 & 11.84 & 11.90 &  5.58\\
%3 & 9.22 & 8.31 & 8.16 & 5.83 & 16.14 & 14.50 & 13.81 &  8.60\\
%4 & 10.66 & 9.09 & 3.91 & 2.11 & 15.09 & 13.07 & 12.32 &  8.35\\
1 & $-3.863$ & 6.990 & 5.144 & 3.989 & $-3.936$ & 16.442 & 5.899 & $-0.316$ \\
2 & $-0.489$ & 11.146 & 2.235 & 1.483 & 4.874 & 20.166 & 5.796 & 6.480 \\
3 & 2.630 & 9.434 & 1.731 & 4.236 & 8.375 & 17.964 & 6.967 & 9.251 \\
4 & $-0.159$ & 7.783 & 1.671 & 3.473 & 3.846 & 16.476 & 4.255 & 8.466 \\
\hline
	\end{tabular} 
\end{table}

\subsection{Study 4: Incestuous sibs}
 \label{sec:incest}
Our remaining examples consider incestuous relationships in two person DNA mixtures of unknown or partly known contributors, in cases of possibly incestuous relationships. This section concerns incest between sibs.

\begin{figure}[htbp]
	\centering
	\includegraphics{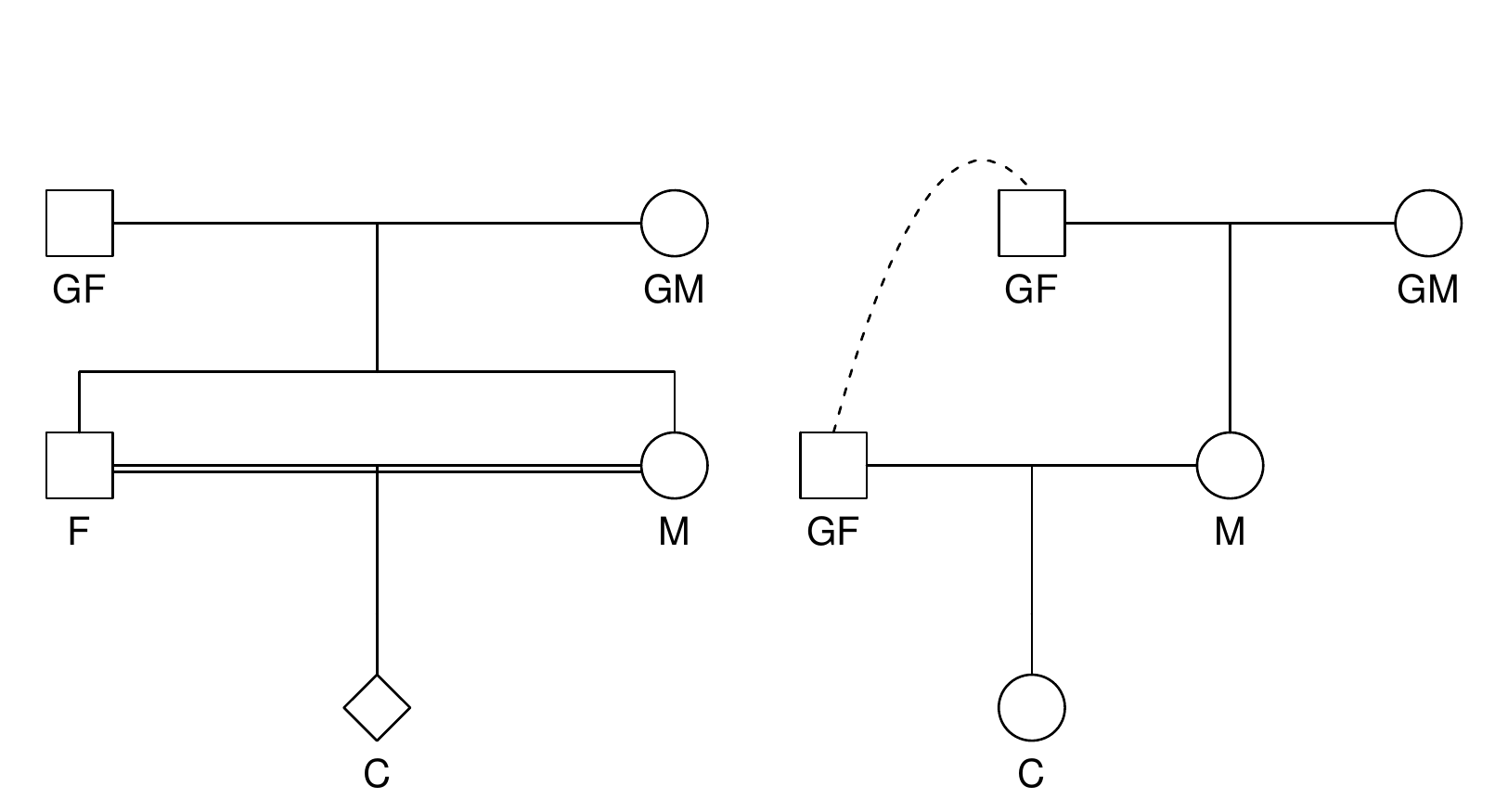}
	\caption{Studies 4 and 5: Pedigrees for ((a), left) incestuous sibs; ((b), right) father--child incest}
	\label{fig:incestpeds}
\end{figure}

The setup we consider is of a father/mother/child trio, and a mixture where the contributors are the mother and child. Cases like this occur when a  mother-foetus mixture is found and we wish to test the paternity. The hypotheses entertained are $\Hp:$ the father and mother are siblings, as in the pedigree in Figure \ref{fig:incestpeds}(a) and $\H_0:$ the father and mother are unrelated. The EPG data are simulated under $\Hp$ in this experiment, and with 200 cells from the mother and 100 from the child.

Table \ref{tab:SibsPat} shows the IBD pattern distribution of the three genotypes under $\Hp$.

\begin{table}[htb]
	\centering
	\caption{Study 4: IBD pattern distribution for the incestuous case where the father F of the child C is the brother of the mother M.}
	\label{tab:SibsPat}
	\vspace{2mm}
	\begin{tabular}{r|rrrrrr}
	\hline
		pr & \multicolumn{2}{c}{F} &  \multicolumn{2}{c}{M}&\multicolumn{2}{c}{C} \\
		\hline
		0.125 &          1 &          2 &          1 &          2 &          1 &          1 \\
		0.125 &          1 &          2 &          1 &          2 &          1 &          2 \\
		0.125 &          1 &          2 &          1 &          3 &          1 &          1 \\
		0.125 &          1 &          2 &          1 &          3 &          1 &          2 \\
		0.125 &          1 &          2 &          1 &          3 &          1 &          3 \\
		0.125 &          1 &          2 &          1 &          3 &          2 &          3 \\
		0.25 &          1 &          2 &          3 &          4 &          1 &          3 \\
		\hline
	\end{tabular}  
\end{table}

\begin{figure}[htbp]
	\begin{center}
		\centering
		\includegraphics{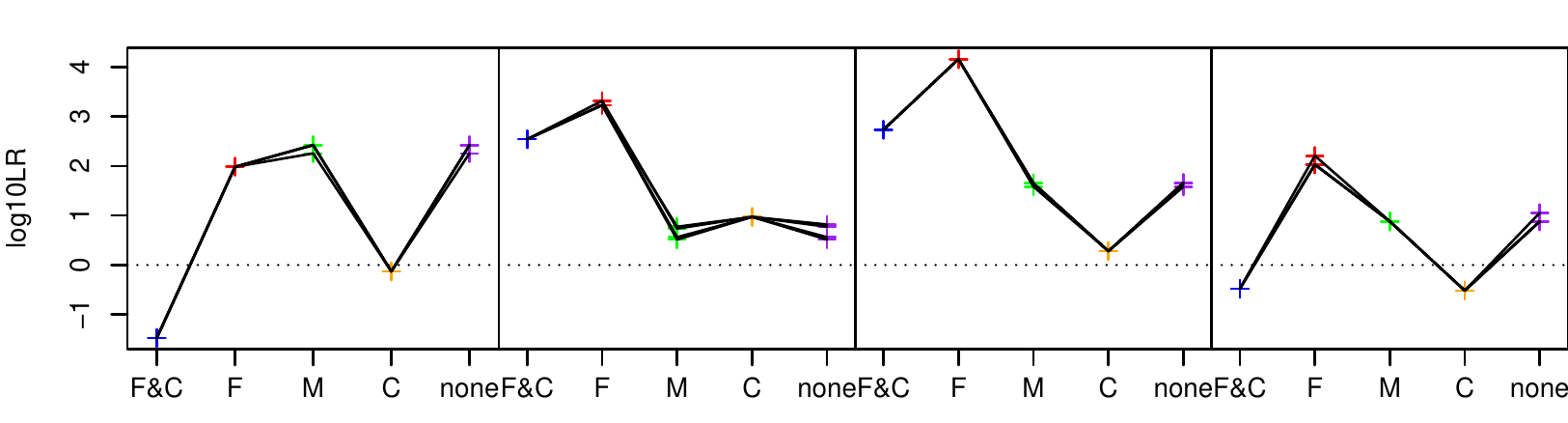}
				\caption{Study 4: $\log_{10} \LR$s for the incestuous sibs example. The panels represent the replicated genotype profiles. In each panel, variation in $\log_{10} \LR$s is shown across replicated EPGs, colour-coded by which actors are genotyped: dark red,  red,  green, orange and purple for F and C, F, M, C and none, respectively.}
		\label{fig:incestsibs}
	\end{center}
\end{figure}

The results for testing whether there was incest or not are shown in Table \ref{tab:incestSibs} and Figure \ref{fig:incestsibs}, the table giving medians over the replicates, the figure showing the variation across $4 \times 4$ replicates. In this example, there is very little variation across EPGs with genotype profiles.

Some of the dependency visible here on which actors are genotyped may seem counter-intuitive. For example, why does typing both Father and Child give apparently less clear evidence of incest than typing either one of Father and Child or neither separately, and indeed in some replicates give evidence against incest? Recall from the beginning of Section 7 that in all of our simulation studies, the focus of interest is on the $\log_{10} \LR$ from the peak height data \emph{given} the stated available genotypes. Taking the peak height data and genotype data \emph{together} (not shown here), as expected, removes the apparent paradox. 

A more careful study of the conditional dependencies in this example reveals that the peak heights convey no information about incest if the Mother and one or other or both of the Father and Child are typed.
\begin{table}[htb]
	\centering
	\caption{Study 4: $\log_{10} \LR$ for testing whether there was incest or not, medians over $4 \times 4$ replicate data sets.}
	\label{tab:incestSibs}
		\begin{tabular}{l|ccccc}
		\hline
		& \multicolumn{5}{c}{Typed actors} \\
			&  F \& C &F &M &C & 	none\\
		\hline
		median $\log_{10} \LR$ & 1.030 & 2.718 & 1.233 & 0.076 & 1.318  \\
		\hline
	\end{tabular}  
\end{table}
\subsection{Study 5: Incest and rape}

Here we consider the horrendous scenario where a child has been raped, and we have a mixed trace from her vagina. The suspected culprit is her maternal grandfather, and additionally there is some suspicion that the grandfather is also her father, that is, that she is the offspring of an incestuous relationship between her mother and maternal grandfather. We assume the child is the major contributor to the mixture, and that there is one other contributor. The pedigree for the case of incest is shown in Figure \ref{fig:incestpeds}(b), and this pedigree is assumed in simulating our EPG data for this study. The EPG data are simulated with 200 cells from the child and 100 from the rapist.
In all cases we do not question  that M is the mother of C, and that GF is the father of M. Table \ref{tab:incestPat} shows the IBD pattern distribution of the three genotypes for an incestuous family where the maternal grandfather \texttt{GF} of a child \texttt{C} is the father of the mother \texttt{M}. 

\begin{table}[htb]
	\centering
	\caption{Study 5: IBD pattern distribution for the incestuous case where the grandfather GF of a child C is the father of the mother M.}
	\label{tab:incestPat}
	\vspace{2mm}
	\begin{tabular}{r|rrrrrr}
	\hline
		pr &        \multicolumn{2}{c}{GF}   &  \multicolumn{2}{c}{M}   & \multicolumn{2}{c}{C}  \\
		\hline
		0.25 &          1 &          2 &          1 &          3 &          1 &          1 \\
		0.25 &          1 &          2 &          1 &          3 &          1 &          2 \\
		0.25 &          1 &          2 &          1 &          3 &          1 &          3 \\
		0.25 &          1 &          2 &          1 &          3 &          2 &          3 \\
		\hline
	\end{tabular}  
\end{table}
  
We consider various possibilities for which actors are separately genotyped: either both M and C, M only, C only, none. As before, our experiments for the above scenarios are based on 4 replicated genotypes by 4 replicated EPGs.
 
\begin{figure}[htbp]
	\begin{center}
		\centering
		\resizebox{18cm}{!}{\includegraphics{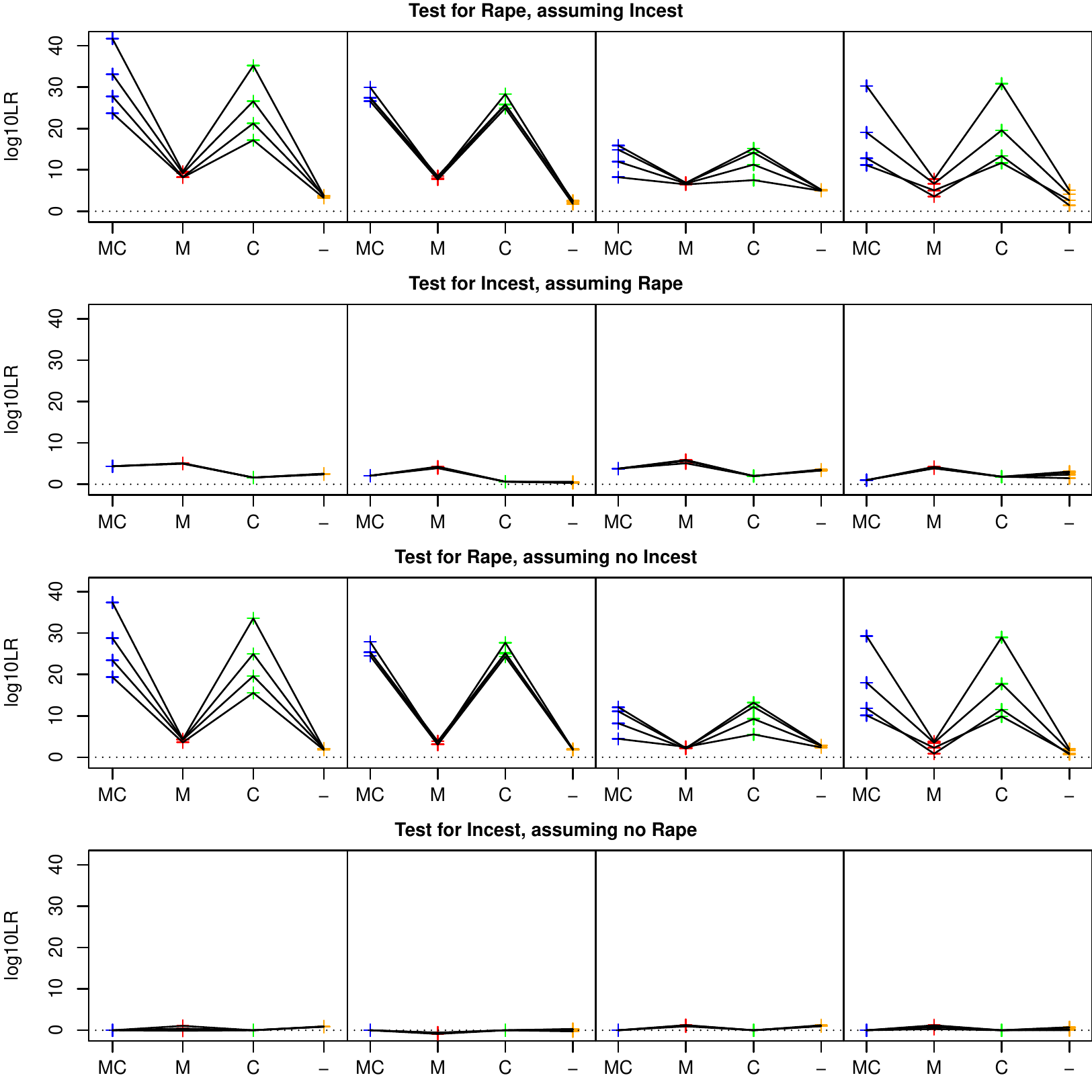}}
				\caption{Study 5: Variation in the $\log_{10} \LR$s across replicate genotype profiles and EPGs for the incest and rape example.
The rows of the figure correspond to the various tests, and the columns to the 4 replicate genotype profiles. Within each panel $\log_{10} \LR$ is plotted against which actors are genotyped (colour-coded: red, red, green, and orange for mother and child, mother alone, child alone, or no one, respectively). Lines join values corresponding to the same EPG replicate.}
		\label{fig:rapeIncest}
	\end{center}
\end{figure}

For brevity, in describing this study we will use the term \textit{rape} to mean that the GF is the 2nd contributor to the mixture, and 
\textit{incest} to mean that GF is father of C.   We wish to examine whether it is possible from the DNA mixture and any typed genotypes to distinguish the possibilities of incest and/or rape.
Table  \ref{tab:incestFC1} reports the $\log_{10} \LR$s for each of (i) rape  assuming incest, (ii) incest assuming rape, (iii) rape  assuming no incest, and (iv) incest assuming no rape. 
Figure \ref{fig:rapeIncest} shows the variation in the $\log_{10} \LR$s by replicate. 
 
%\todo{NOTE: Anomaly in \ref{tab:incestFC1}  as in first row when none are typed we get a larger value than when M and C are typed. Also be consistent with ordering of 4 cases. }

Note that for the test of incest assuming no rape, when the child's genotype is known, a conditional independence argument confirms that the $\log_{10} \LR$ is identically 0.
 
 \begin{table}[htb]
 	\centering
 	\caption{Study 5: $\log_{10} \LR$s for the  tests (i)-(iv) over 4 replicated genotypes by 4 replicated epgs with the set of typed actors given in (b).}
 	\label{tab:incestFC1}
	\vspace{2mm}
 \begin{tabular}{l|rrrr}
\hline
 	& \multicolumn{4}{c}{typed} \\
 	 scenarios                        & both M and C &         M  &          C &       none \\
 	\hline
% 	rape assuming incest &                   22.21 &   8.35  & 20.72  &  4.33 \\
% 	incest assuming rape &                  2.97  &  4.22  &  1.86  &  2.35 \\
% 	rape assuming no incest &                   19.79  &  4.10 &  18.69  &  2.56 \\
% 	incest assuming no rape &                   0.00 &  $-0.16$  &  0.00 &  $-0.05$ \\
  rape assuming incest &  25.145 & 7.713 & 20.403 & 3.679 \\
 	incest assuming rape &  2.916 & 4.608 & 1.721 & 2.506 \\
 	rape assuming no incest & 21.404 & 3.111 & 18.682 & 2.002 \\
 	incest assuming no rape & 0 & 0.574 & 0 & 0.787 \\
	\hline
 \end{tabular}  
  \end{table}

We can see that there tends to be a stronger signal for rape than for incest, and also that when we have the additional information on the child's genotype the $\log_{10} \LR$ becomes much larger. Having only the mother's genotype does not make a substantial difference. 
 
\section{Real case applications}

\label{sec:real case}
\subsection{Case 1: Identification of a missing person}

Here we analyse a real case related to a missing male, provided by the DNA Laboratory, Comisar\'{i}a General de Polic\'{i}a Cient\'{i}fica of Madrid. We refer to  \textcite{green-mortera-prieto} for  details of the case analysis.  Here we will revisit some of the results, and also compare them with an  analysis made with different software.  The full anonymised data together with the \texttt{R} scripts to compute the results are given in the  webpage\footnote{\texttt{https://petergreenweb.wordpress.com/example-1-data-code-and-output/}
}. The data are anonymised to avoid serious privacy and confidentiality concerns.

In this  case, only a daughter of the missing male was available to donate a DNA sample. A biological sample was collected from  a toothbrush, presumably used by the missing person. DNA  was recovered and analysed on 21 STR markers, plus three sex related markers, included in GlobalFiler\textsuperscript{TM} PCR amplification kit (ThermoFisher). The reference sample from the daughter of the missing male was also genotyped with the same kit. A two-person DNA mixture was detected on the toothbrush. We denote the two distinct unknown contributors by $U_1$ for the major contributor and $U_2$ for the minor. An excerpt of the data is shown in Table \ref{tab:data}, showing the alleles and peak heights in the  DNA mixture found on the toothbrush $T$. The DNA profile of the daughter, denoted by D, is also shown. The sex related markers (not shown) indicated that the mixture was most probably from a male and a female donor.  

\begin{table}[htb]
	\centering
	\caption{Case 1: An excerpt of the anonymised data from the toothbrush $T$, showing the markers, alleles and relative peak heights together with  the daughter's genotype D.}
	\label{tab:data}
	\vspace{2mm}
	\begin{tabular}{l|rrr}
		\hline  	
		&       alleles    &toothbrush             &   \\
		markers &   in mixture  & peak height &    D \\
		\hline
		Marker 6   & 19 &        264  &       19  \\    
		& 21 &        3664 &       21  \\
		Marker 7   & 13  &       1152 &                   \\
		& 14 &         126 &                 \\
		&15  &         941 &   15    \\
		Marker 14 & 13 &       5158  &   13  \\
		& 15 &        304 &       15   \\
		Marker 20 & 13 &       3218 &                 \\
		& 17 &       3550 &       17  \\
		\hline
	\end{tabular}  
\end{table}

Here we analyse the  DNA mixture found on the toothbrush $T$  presumably used by the missing person. 
We assume known  allele frequencies and adopt a threshold of 50 rfus (relative florescent units).

In the preliminary analysis of the data we found that the estimated proportion of DNA for the  major contributor $U_1$, $\phi_{U_1}=0.93$ is large. We also computed
the $\log_{10} LR$ for testing  $\Hp$: D is the child of $U_1$ (and similarly for  $U_2$)  {\emph{vs.}} 
$\H_0$: $U_1$ and $U_2$   are assumed drawn at random from the gene pool.  The $\log_{10} \LR=10.97$ is large, pointing to  $U_1$ being a parent of D but it is also substantial, $\log_{10} \LR=4.53$, for the hypothesis that the minor contributor $U_2$ is a parent of D. We then  questioned whether $U_1$ and $U_2$ were possibly related.  The negative $\log_{10} \LR$s for comparing $\Hp$:  $U_1$ has the relationship to $U_2$,   \textit{vs.} $\H_0$: $U_1$ and $U_2$  are unrelated, given in Table \ref{tab:related}, show that they are most likely to be unrelated and  thus are likely to be the parents of D or close relatives thereof. Furthermore, the likelihood that D  was one of the contributors is practically zero. 

\begin{table}[htb]
	\centering
	\caption{Case 1:  The $\log_{10} \LR$ for $\Hp$: the two contributors to the mixture are related as specified \textit{vs.} $\H_0$: the two contributors are unrelated.}
	\label{tab:related}
	\vspace{2mm}
	\begin{tabular}{r|r}
		\hline
		Relationship between $U_1$  &  $T$  \\
		and $U_2$ under $\Hp$          & \multicolumn{1}{c}{$\log_{10} \LR$}     \\
		\hline
		monozygotic twins &  $- \infty$      \\
		parent-child &  $- \infty$    \\
		sibs &  $-2.14$  \\
		double first cousins & $-0.51$  \\
		half-sibs & $-0.37$  \\
		first cousins & $-0.10$   \\
		half-cousins & $-0.034$\\
		\hline
	\end{tabular}  
\end{table}

We now examine the joint relationships between the mixture contributors and the typed daughter D, which clarifies the role of D in validating the mixture profile. Table \ref{tab:Trel} shows the  $\log_{10} \LR$ for item $T$  for several hypotheses $\Hp$ concerning different relationships among $U_1$, $U_2$ and D,  \textit{vs.} the null hypothesis that these individuals are all unrelated. The values of the $\log_{10} \LR$ show that there is strong evidence that the two contributors to item $T$ are the missing father of D and D's mother, or at least very close relatives of them. Comparing the first 4 rows of Table \ref{tab:Trel} confirms that the most likely single possibility is that they are indeed the mother and father. All values in the Table remain unchanged if the sexes of all contributors are reversed; we choose to identify them in the way shown because inference (not shown) also including the Amelogenin locus indicates that is is most likely that the major contributor $U_1$ is female.

If there is interest in comparing two of the models displayed in Table \ref{tab:Trel}, the appropriate $\log_{10} \LR$ is simply obtained by calculating the difference between the values shown. For example, comparing the first row and the fifth, $17.935-10.974 = 6.961$ gives the weight of evidence that $U_2$ is the father of D, given that it is already assumed that $U_1$ is the mother of D. There are too many different such comparisons that can be made to list them all here.

Some of the specific relationships examined in Table \ref{tab:Trel} are speculative, but might be of interest in cases where a home is shared by an extended family.

\begin{table}[htb]
	\centering
	\caption{Case 1: For item $T$, $\log_{10} \LR$ for several hypotheses $\Hp$ concerning different relationships among $U_1$, $U_2$ and D,  \textit{vs.} $\H_0$: $U_1$, $U_2$ and D are all unrelated. The results in the lower half of the table can be used as baselines for comparison for those in the upper half. All $\log_{10} \LR$ remain unchanged if the sexes of $U_1$ and $U_2$  are switched.}
	\label{tab:Trel}
	\vspace{2mm}
	\begin{tabular}{l|r}
		\hline
		\multicolumn{1}{c|}{$\H_p$} &     $\log_{10} \LR$        \\
		\hline
		$U_1$ mother of D and $U_2$ father of D  &          17.935 \\
		$U_1$ maternal aunt of D and $U_2$ father of D &          14.028 \\
		$U_1$ mother of D and $U_2$ paternal uncle of D &         15.579\\
		$U_1$ maternal aunt of D and $U_2$ paternal uncle of D &        11.763\\
		\hline
		$U_1$ mother of D and $U_2$ unrelated & 10.974    \\
		$U_1$ maternal aunt of D and $U_2$ unrelated & 7.452    \\
		$U_1$ unrelated and $U_2$ father of D &  4.530   \\
		$U_1$ unrelated and $U_2$ paternal uncle of D  &  2.796   \\
		\hline
	\end{tabular}  
\end{table}

\subsubsection{Case 1, comparison with {\tt relMix}}

\begin{table}[htb]
	\centering
	\caption{Case 1: Marker-wise $\LR$ and overall $\log_{10}\LR$ for item $T$, using \texttt{relMix} and
		\texttt{KinMix} with and  without peak height information, for testing whether in $T$,  $\H_p$: $U_1$  is the father of D \textit{vs.} $\H_0$: $U_1$ and $U_2$ are random members of the population. The partial $\log_{10}\LR$ values exclude markers 8 and 10.}
	\label{tab:RelKin2}
	\vspace{2mm}
	\begin{tabular}{l|rrr}
		\hline
		marker	&     \texttt{relMix} &     \texttt{KinMix}  &     \texttt{KinMix}  \\
		&                     &     w/o peak heights &  with peak heights \\
		\hline
		1 &       3.02 &       3.11 &       4.84 \\
		2 &       7.18 &       7.77 &      10.92 \\
		3 &       9.23 &       9.81 &      12.89 \\
		4 &       3.54 &       3.61 &       3.72 \\
		5 &       4.75 &       4.93 &       5.16 \\		
		6 &       2.55 &       2.58 &       3.34 \\
		7 &       1.08 &       1.07 &       1.59 \\
	    8 &        NaN &       5.53 &       5.07 \\
		9&       1.26 &       1.18 &       1.62 \\
		10 &        NaN &       2.34 &       2.15 \\
	    11 &       6.30 &       6.35 &      10.92 \\
	    12 &       1.71 &       1.71 &       1.99 \\
		13 &       2.25 &       2.26 &       2.27 \\
		14 &       2.09 &       2.12 &       1.51 \\
		15 &       2.39 &       2.46 &       2.54 \\
		16 &       5.02 &       5.24 &       6.17 \\
		17 &       1.89 &       1.74 &       2.39 \\
		18 &       3.48 &       3.52 &       6.77 \\
		19 &       3.01 &       3.11 &       3.30 \\
		20 &       1.60 &       1.49 &       1.23 \\
		\hline
		partial  	$\log_{10}\LR$&       8.35 &       8.42 &       9.94 \\
		computation times (s)      &   1314 &    2.30     &  2.32  \\    
		%computation times (s) P     &   1070 &    2.06     &  2.04  \\    
		overall 	$\log_{10}\LR$ &       & 9.53 &    10.97  \\
		computation times (s)      &    &   3.02      &  3.09 \\    
		%computation times (s)  P    &    &   2.64     &  2.63 \\    
		\hline
	\end{tabular}  
\end{table}
The \texttt{R} package \texttt{relMix} \cite{hernandis2019relmix} also analyses DNA mixtures involving relatives, but is based only on allele presence, not considering the peak heights when modelling the DNA mixture as does \texttt{KinMix}. Based on the toothbrush sample, here we draw a comparison between results when using \texttt{KinMix} with and without the peak height data, and using \texttt{relMix}. 

Table \ref{tab:RelKin2} presents a  marker-wise comparison between the likelihood  \LR\ and the overall $\log_{10} \LR$, for comparing  $\Hp$: $U_1$ is the father of D {\emph{vs.}} 
$\H_0$ $U_1$ and $U_2$  are assumed drawn at random from the gene pool.
Note that on comparing columns 2 and 3 to column 4 in Table \ref{tab:RelKin2}, for only 2 markers out of 20 does using the peak heights yield a smaller $\log_{10} \LR$ than using allele-presence data alone. The results obtained with  \texttt{relMix} and \texttt{KinMix} when not including  the peak height information are quite similar, although based on quite different models.  

We found that \texttt{relMix} was not able to compute the \LR\ for markers 8 and 10; these are the markers for which there are more than 16 different alleles in the allele frequency database,. We indicate this in Table \ref{tab:RelKin2} by NaN.  When  using only those markers that  \texttt{relMix} is able to compute, the partial $\log_{10}\LR$ obtained  with \texttt{KinMix} with peak height information, is $9.94$, is substantially bigger than that obtained without peak height information ($8.35$ with \texttt{relMix}, $8.42$ with \texttt{KinMix}).  

Using \texttt{KinMix} on all the markers, the overall $\log_{10} \LR$ is $10.97$ with peak heights, compared with the result $9.53$ when not using the peak height information, a $\LR$ 27.5 times smaller.

The time to do the computations with \texttt{relMix} is considerably longer; it takes 1,314 seconds (22 minutes) compared to 2.30 seconds  for \texttt{KinMix} without peak height information, and the 2.32 seconds using peak heights. (These times were obtained with an i7-7600U processor clocked at 2.80GHz.) 
This discrepancy between computational times is likely to be due to 
differences in the data structures used to represent relationships in DNA mixtures.

\subsection{Case 2: Rape case}

Here we analyse another real case  from the DNA Laboratory, Comisar\'{i}a General de Polic\'{i}a Cient\'{i}fica of Madrid, concerning a sexual assault.
The victim stated that 2 individuals raped her, both using condoms. The two condoms were found and two DNA mixtures, EPG1 and EPG2, were detected and analysed on 22 markers including Amelogenin (for gender identification). The two DNA mixtures were each 
compatible with the victim's profile  and with that of an unknown male.
An excerpt of the data on two DNA mixtures EPG1, EPG2 showing the markers, alleles, peak heights and victim's genotype is shown in Table \ref{tab:rapedata}.

\begin{table}[htb]
	\centering
	\caption{Case 2: An excerpt of the data on two DNA mixtures EPG1, EPG2 showing the markers, alleles, peak heights and victim's genotype.}
	\label{tab:rapedata}
	\vspace{2mm}
	\begin{tabular}{r|rrrr}
		\hline
		&            &       EPG1 &       EPG2 &            \\
		marker     &     allele &     height &     height &     victim \\
		\hline
		CSF1PO &         10 &       1449 &        173 &         10 \\
		&         11 &        133 &        129 &            \\
		D10S1248 &         13 &       6380 &       1527 &         13 \\
		&         14 &       1012 &        139 &            \\
		D12S391 &         19 &        172 &            &            \\
		&         20 &       1001 &        152 &         20 \\
		&         21 &       1193 &         73 &         21 \\
		D18S51 &         12 &            &         88 &            \\
		&         14 &        363 &            &            \\
		&         15 &        791 &        271 &         15 \\
		&         16 &       1469 &        232 &         16 \\
		&         19 &        461 &            &            \\
		\hline
	\end{tabular}  
\end{table}

The police wished to know  if the unknown male in EPG1 and that in EPG2  could belong to 2 brothers. We also tested whether the two unknown males have a relationship  $R=$\{father--son, brothers, half-brothers, cousins, or are identical (the same individual or a hypothetical monozygotic twin)\}.

In order to  deal with problems like this  we consider having the victim, $v$, and  two unknown contributor $U_1$ and $U_2$ in EPG1  and   EPG2,  and assume that the proportion contributed to EPG1 by $U_2$ is $\phi_{U_2}=0$  and the proportion contributed to EPG2 by $U_1$  is $\phi_{U_1}=0$. The estimated mixture parameters for EPG1 and EPG2, under the above assumption, are given in Table \ref{tab:rapepar}.

\begin{table}[htb]
	\centering
	\caption{Case 2: Estimated parameters for EPG1 and EPG2 assuming they contain DNA from the victim $v$ and two unknown contributors $U_1$ and $U_2$. We assume that the proportion contributed to EPG1 by $U_2$ is zero, $\phi_{U_2}=0$,  and the proportion contributed to EPG2 by $U_1$  is zero,  $\phi_{U_1}=0$. }
	\label{tab:rapepar}
	\begin{tabular}{rrrrrrr}
		\hline
		&        $\rho$ &     $\eta$ &        $\xi$ &    $\phi_{U_1}$ &    $\phi_{U_2}$ &      $\phi_{v }$ \\
		\hline
		EPG1 &       4.786 &  432.7    &          0 &      0.179 &     0 &      0.821 \\
		EPG2 &        1.996 &  289.9&       0.0303 &          0 &      0.336 &      0.664 \\
	\end{tabular}
\end{table}

\begin{table}[htb]
	\centering
	\caption{Case 2: $\log_{10} \LR$ for testing $\Hp: \; U_{1} \;\; \mbox{in EPG1 and} \; \; U_{2}$ in EPG2 have the specified relationship \textit{vs.} $\H_0: \; U_{1} \;\; \mbox{in EPG1 and} \; \; U_{2}$ in EPG2 are unrelated. In both $\Hp$ and $\H_0$ we assume that the victim is a contributor to both EPG1 and EPG2. }
	\label{tab:rapeLR}
	\vspace{2mm}
	\begin{tabular}{rrrrrr}
		\hline
		&   identity & father--son &       brothers &  half-brothers &    cousins \\
		\hline
		relationship &       2.36 &        2.60 &       2.48 &       2.56 &       1.84 \\
		\hline
	\end{tabular}  
\end{table}

Table \ref{tab:rapeLR} gives the $\log_{10} \LR$ for testing $\Hp: \; U_{1} \;\; \mbox{in EPG1 and} \; \; U_{2}$ in EPG2 have the specified relationship \textit{vs.} $\H_0:$ all contributors unrelated, except for the victim's presence in both.  
The  $\log_{10} \LR$ points to there  being a relationship between $U_1$ and $U_2$. In particular, the $\log_{10} \LR =2.48$ that they are brothers rather than unrelated. 

In subsequent developments of the case two brothers were arrested and DNA reference samples were collected from them.  We  compared the prosecution hypothesis $\Hp: v $ and brother 1 contributed to EPG1 \textit{vs.} $\H_0: v $ and an unknown, selected at random from the database,  contributed to EPG1. The resulting $\log_{10} \LR= 22.28$  for $\Hp$ \textit{vs.} $\H_0$ is 
is highly incriminating for brother 1. Similarly, when testing  the prosecution hypothesis $\Hp: v $ and brother 2 contributed to EPG2 \textit{vs.} $\H_0: v $ and an unknown, selected at random from the database,  contributed to EPG2, the $\log_{10} \LR= 24.14$  for $\Hp$ \textit{vs.} $\H_0$  
is highly incriminating also for brother 2.

\section{Discussion}

We have shown that the IBD pattern distribution for a collection of related individuals, which extends Jacquard's concept of coefficient of identity by descent beyond pairwise relationships, is an invaluable approach both to encoding relationships and to BN computations for DNA mixture analysis involving family relationships. Implementation of these ideas in the package \texttt{KinMix}, extending \texttt{DNAmixtures}, provides a convenient, powerful and flexible means for delivering the computations needed for DNA mixture analysis, using peak heights, involving arbitrarily complex relationships.

In this paper, we have not paid attention to the possibility of mutation, which can be important, for example in paternity cases, where a putative father can be excluded because his genotype profile is incompatible, even though a single mutation could make it compatible. We have experimental additions to \texttt{KinMix}, that compute likelihood ratios allowing for mutation, in various scenarios involving very close relationships. We aim to provide a comprehensive treatment of mutation in subsequent work.

%In further work, we intend to provide extensions, such as handling heterogeneous sub-populations, and to investigate the possibility of including mutation, and additional PCR artefacts, into this framework.

\subsection*{Acknowledgements}

We thank Amke Caliebe, Thore Egeland and Michael Nothnagel for organising an excellent workshop on Advanced Statistical and Stochastic Methods in Forensic
Genetics, in Cologne in August 2018, and acknowledge fruitful conversations with Marjan Sjerps, Nuala Sheehan, Torben Tvedebrink, and Magnus Dehli Vigeland at and after the meeting, and also helpful correspondence with Elizabeth Thompson, and comments from Phil Dawid.
Oskar Hansson was very helpful in discussion about using his package \texttt{pcrsim}. We also thank Lourdes Prieto,
Comisar\'{i}a General de Polic\'{i}a Cient\'{i}fica, DNA Laboratory, Madrid, Spain for providing the data for the real case applications.

Finally, we are indebted to the referees for their careful reviews, which have led to a much improved presentation of these ideas.

\newpage

\section*{Supplementary information}

\section*{Supplementary section 1: Algorithms in pseudo-code}
See Algorithm \ref{generative} for the pseudocode for the generative model, that is for generating genotypes from the constructive model of Section 3.2. The algorithm for the construction of the CPTs for the BNs (Sections 3.3 and 3.4)
is essentially the same except that random variables are replaced by their probability distributions in table form, as in Algorithm \ref{cptscode}.

The notation used is as follows. The input variables are a coding of some of the information illustrated in Table 4: $\pi$ is one of the permuted patterns that can generate the typed individual genotypes (in this example, 5 in number); $ncontr$ is the number of mixture contributors whose genotypes's conditional distributions we are modelling (in the example in Table 4, two: {\tt Fgt} and {\tt Mgt}), these are numbered $1=1,2,\ldots,ncontr$ in the pseudocode; $g=1,2$ indexes an individual's paternal and maternal genes; $draw_{\pi ig}$ is a boolean variable, true if the $g$th gene of individual $i$ has to be drawn from the gene pool under permuted pattern $\pi$, false if it is determined by the typed genotypes; $par_{\pi ig}$ is an integer parameter, the allele index if $draw_{\pi ig}$ is false, a running count of genes to be drawn under this pattern if $draw_{\pi ig}$ is true; $p(\pi)$ is the probability distribution of $\pi$ given the observed genotypes (in this example, given {\tt Cgt} and {\tt GFgt}); $ndraws$ is the maximum over $\pi$ of the number of draws needed under pattern $\pi$; $(q_a)$ are the allele frequencies and
$q^\star_a=q_a/\sum_{b=1}^{a-1} q_b$. As in our example, so in general, all of these input values are easily determined from the specified IBD pattern distribution and the allele frequencies.

The output variables are the genotype allele count arrays $(n_{ia})$ for the mixture contributors $\{i\}$, whose distribution conditions on the observed typed genotypes of their relatives, together with the latent variables $\pi$, the permuted pattern, and $(m_{ja})$, the allele count arrays for the gene pool draws. 

\begin{algorithm}[h]
\algrenewcommand\algorithmicrequire{\textbf{Input:}}
\algrenewcommand\algorithmicensure{\textbf{Output:}}
\caption{The generative model}\label{generative}
\begin{algorithmic}[1]
\Require $ndraws,ncontr,(draw_{\pi ig}),(par_{\pi ig}),(q_a),p(\pi)$
\Ensure Sample values from the joint distribution of nodes $\pi,(m_{ja}),(n_{ia})$
\For{$j \gets 1,2,\ldots,ndraws$} \Comment Draws from the gene pool
\State $m_{j1}\gets Bernoulli(q_1)$
\State $T_{j1} \gets m_{j1}$
\For{$a \gets 2,3,\ldots,A-1$}
\State $m_{ja}\gets Bernoulli((1-T_{j,a-1})q^\star_a)$
\State $T_{ja}\gets T_{j,a-1}+m_{ja}$
\EndFor
\State $m_{jA}\gets (1-T_{j,A-1})$
\EndFor
\State Draw $\pi$ w.p. $p(\pi)$ \Comment Drawing from pattern distribution
\For{$i \gets 1,2,\ldots,ncontr$}
\For{$a \gets 1,2,\ldots,A$}
\For{$g \gets 1,2$} 
\If{$draw_{\pi ig}$}
 $j \gets par_{\pi ig}$;
 $h_g\gets m_{ja}$
\Else
    $\;h_g\gets I[par_{\pi ig}=a]$
\EndIf
\EndFor
\State $n_{ia} \gets h_1+h_2$ 
\EndFor
\EndFor
\end{algorithmic}
\end{algorithm}

\begin{algorithm}[h]
\algrenewcommand\algorithmicrequire{\textbf{Input:}}
\algrenewcommand\algorithmicensure{\textbf{Output:}}
\caption{Constructing CPTs}\label{cptscode}
\begin{algorithmic}[1]
\Require $ndraws,ncontr,(draw_{\pi ig}),(par_{\pi ig}),(q_a),p(\pi)$
\Ensure CPTs for a BN with nodes $\pi,(m_{ja}),(n_{ia})$
\For{$j \gets 1,2,\ldots,ndraws$} \Comment Draws from the gene pool
\State $p(m_{j1}) \gets q_1I[m_{j1}=1]+(1-q_1)I[m_{j1}=0]$
\State $p(T_{j1}) \gets I[T_{j1}=m_{j1}]$
\For{$a \gets 2,3,\ldots,A-1$}
\State $p(m_{ja}=1) \gets q^\star_aI[T_{j,a-1}=0]$ ; $p(m_{ja}=0)=1-p(m_{ja}=1)$
\State $p(T_{ja})\gets I[T_{ja}=T_{j,a-1}+m_{ja}]$
\EndFor
\State $p(m_{jA})\gets I[m_{jA}=(1-T_{j,A-1})]$
\EndFor
\ForAll{$\pi$} \Comment Looping over pattern distribution
\For{$i \gets 1,2,\ldots,ncontr$}
\For{$a \gets 1,2,\ldots,A$}
\For{$g \gets 1,2$} 
\If{$draw_{\pi ig}$}
 $j \gets par_{\pi ig}$;
 $h_g\gets m_{ja}$
\Else
    $\;h_g\gets I[par_{\pi ig}=a]$
\EndIf
\EndFor
\State $p(n_{ia}|\pi,(m_{ja})) \gets I[n_{ia}=h_1+h_2]$ 
\EndFor
\EndFor
\EndFor
\end{algorithmic}
\end{algorithm}

\section*{Supplementary section 2: Uncertainty in allele frequencies in complex problems involving relationships}

Let $T$ be the typed genotypes, $C$ the genotypes of the contributors to the mixture, $M$ the meiosis pattern, $f$ the gene pool draws, and $q$ the allele frequencies.

Then we want $P(C|T)$.

Clearly
%\begin{align}
$$ 
P(C|T) = \frac{P(C,T)}{P(T)} = \frac{\int \sum P(C,T|M,q,f) p(M) p(f|q) p(q) dq}{\int \sum P(T|M,q,f) p(M) p(f|q) p(q) dq}
$$ %\\
%&= \frac{\int \sum p(C|M,q,f_C) p_{f_C} P(T|M,q,f_T) p_M p_{f_T} p(q) dq}{\int \sum P(T|M,q,f_T) p_M p_{f_T} p(q) dq}.
%\end{align}
$T$ is a deterministic function of $M$ and of a subvector of $f$ we call $f_{T,M}$, so the denominator is
$$
P(T) = \int \sum P(T|M,q,f) p(M) p(f|q) p(q) dq = \sum_M \sum_{f_{T,M}} P(T|M,f_{T,M}) p(M) \int p(f_{T,M}|q) p(q)dq,
$$
in which $P(T|M,f_{T,M})$ is actually an indicator function.

$C$ is a deterministic function of $M$, $f_{T,M}$ and additional gene pool draws $f_{C|T,M}$, so 
$$
P(C,T) = \sum_M \sum_{f_{T,M}} \sum_{f_{C|T,M}} P(C|M,f_{T,M},f_{C|T,M}) P(T|M,f_{T,M}) p(M) \int p(f_{T,M}|q)p(f_{C|T,M}|q)p(q)dq
$$

\subsubsection*{Dirichlet--Multinomial}
$p(f_{T,M}|q)$ and $p(f_{C|T,M}|q)$ are polynomials in $q$, and $p(q)$ is a Dirichlet distribution, so $\int p(f_{T,M}|q) p(q)dq$ and $\int p(f_{T,M}|q)p(f_{C|T,M}|q)p(q)dq$ are explicit (expressed in terms of Gamma functions).

Suppose that the allele frequencies $q$ have a Dirichlet prior: $p(q)=\prod_a q_a^{\delta_a-1}\times \Gamma(\sum_a \delta_a)/\prod_a \Gamma(\delta_a)$, and that $f_{T,M}$ consists of $n_a$ copies of allele $a$, $a=1,2,\ldots,A$. Then $p(f_{T,M}) = \prod_a q_a^{n_a}$ and
$$
\int p(f_{T,M}) p(q) dq = \int \prod_a q_a^{\delta_a+n_a-1} dq \times \frac{\Gamma(\sum_a \delta_a)}{\prod_a \Gamma(\delta_a)}
= \frac{\Gamma(\sum_a \delta_a)\prod_a \Gamma(\delta_a+n_a)}{\Gamma(\sum_a (\delta_a+n_a))\prod_a \Gamma(\delta_a)}
=DM(n;\delta),
$$
say. Similarly, if $f_{C|T,M}$ has $n'_a$ copies of allele a, then
$$
\int p(f_{T,M}|q)p(f_{C|T,M}|q)p(q)dq = \frac{\Gamma(\sum_a \delta_a)\prod_a \Gamma(\delta_a+n_a+n'_a)}{\Gamma(\sum_a (\delta_a+n_a+n'_a))\prod_a \Gamma(\delta_a)} = DM(n+n';\delta).
$$
(These Dirichlet--Multinomial distributions are what is evaluated in the P\'olya urn BN.)
\subsubsection*{Computation of $P(C|T)$}

$T$ is observed, so we only need $P(T|M,f_{T,M})$ for a single value of $T$, and therefore $P(T)$ becomes an explicit sum over $M$ and $f_{T,M}$, with only a small number of non-zero terms.

We need $P(C|T)$ for all possible values of $C$, so to compute $P(C|T)$ we use a BN whose nodes include $f_{C|T,M}$ and $M$ ($T$ being now fixed and $f_{T,M}$ a fixed function of $M$). In this BN, if the evidence and joint probability has product $P(C,T)/P(T)$, then the normalising constant will evaluate $P(C|T)$.
\begin{multline*}
P(C,T)/P(T) = \\ P(T)^{-1} \sum_M \sum_{f_{T,M}} \sum_{f_{C|T,M}} P(C|M,f_{T,M},f_{C|T,M}) P(T|M,f_{T,M}) p(M) DM(n+n';\delta)\\
= P(T)^{-1} \sum_M \sum_{f_{T,M}} \sum_{f_{C|T,M}} P(C|M,f_{T,M},f_{C|T,M}) P(T|M,f_{T,M}) p(M) DM(n';n+\delta)DM(n;\delta)
\end{multline*}

Of the terms in this expression, $P(T)$, $\{\delta_a\}$, $\{n_a\}$ are constants, $P(C|M,f_{T,M},f_{C|T,M})$ and $P(T|M,f_{T,M})$ are indicator functions. Write $p^\star(M)=P(T)^{-1} p(M) \sum_{f_{T,M}} P(T|M,f_{T,M})DM(n;\delta)$ (which is explictly available but may not be correctly normalised), then 
$$
P(C,T)/P(T) = \sum_M \sum_{f_{C|T,M}} P(C|M,f_{T,M},f_{C|T,M}) p^\star(M) DM(n';n+\delta).
$$
However, note that $n$ are the allele counts corresponding to $f_{T,M}$, so for fixed $T$ vary with $M$. This means that the P\'olya urn component of the BN will have the node corresponding to $M$ as an additional parent. Alternatively, we have to loop over $M$, running a separate BN for each value (for which $p^\star(M)$ is non-zero), and combining them afterwards.

\section*{Supplementary section 3: Software for interpreting DNA mixtures in the presence of relationships}

\begin{table}[h]
%\clearpage
	\caption{Summary of main software programs available and their characteristics.}
	\label{tab:mix}
	\centering
	%\begin{threeparttable}
		\begin{tabular}{l|c|l|c}
			Software             & Open source & Model     &Relatedness           \\
						\hline
			\DNAmixtures\ (a) { \&}          &   &           &                            \\
			\texttt{KinMix} (b)                 & yes           & Continuous & yes               \\
			\texttt{EuroForMix} (c)          & yes           & Continuous &   $F_{st}$ or $\theta$ correction                \\
			\texttt{LiRa} (d)                &   for internal use & Continuous &  $F_{st}$ or $\theta$ correction               \\
%			&      use only      &                 \\	{\texttt{TrueAllele}}\textsuperscript{\textregistered}{\tnote{8}}              & no & Continuous & no                  \\
			{\texttt{STRmix}}\textsuperscript{\texttrademark} (e)                & no               & Continuous &   stated on website                \\
			\texttt{likeLTD} (f)               & yes           & Semi-continuous   &    $F_{st}$ or $\theta$ correction                \\
			\texttt{relmix} (g) &  yes & Discrete & yes \\
			\texttt{Forensim} (h)             & yes           & Discrete   &      $F_{st}$ or $\theta$    correction         \\ 
%			Lab Retriever{\tnote{6}}               & yes           & Discrete   & no                  \\ 		         
			\hline
		\end{tabular}
	\end{table}
	%	\begin{tablenotes}

Given the problems with subjective interpretation of complex DNA mixtures, a number of researchers have developed computer programs for   ``probabilistic genotyping'' that apply various models to interpret DNA mixtures. Some  use discrete or semi-continuous methods, which
use the allele information possibly in conjunction with probabilities of allelic dropout
and dropin, while other programs  also include the continuous peaks in varying ways, and some also adjust for  relatedness using the co-ancestry coefficient $F_{st}$ or  the $\theta$ correction. Table \ref{tab:mix} presents the current software that, to our knowledge, deals with relationship  in  DNA mixture analysis.  These systems vary in the input data they use,  the modelling assumptions they make, and thus the resulting analysis.  These programs clearly represent a major improvement over purely subjective interpretation.

	\begin{enumerate}
			\item  Graversen, T. (2013). \texttt{DNAmixtures}: Statistical inference for mixed traces of DNA. \texttt{R} package version 0.1-4. \texttt{http://dnamixtures.r-forge.r-project.org/}. It requires {\tt Hugin}.
			 \item  Green, P. J. (2020). \texttt{KinMix}: DNA mixture analysis with related contributors. \texttt{R} package version
		 2.0. \texttt{https://petergreenweb.wordpress.com/kinmix2-0}.
			\item Bleka, O.,  Storvik,  G.,  Gill, P. (2016). \texttt{EuroForMix}: an open source software based on a continuous model to evaluate STR DNA profiles from a mixture of contributors with artefacts, \textit{Forensic Sci. Int. Genet.}, \textbf{21},  35--44. \texttt{http://www.euroformix.com/}
			\item Puch-Solis, R., Rodgers, L,.  Mazumder,  A.,  Pope, S.,  Evett, I.W.,  Curran, J., and Balding, D. (2013). Evaluating forensic DNA profiles using peak heights, allowing for multiple donors, allelelic dropout and stutters, \textit{Forensic Science International: Genetics}, \textbf{7}, 555--63.			 \\
			 \verb+https://cdnmedia.eurofins.com/european-west/media/1418957/+ \\ \verb+lgc_lira_fact_sheet_en_0815_90.pdf+
			\item Taylor, D.,   Bright, J. A.,  Buckleton, J. (2013). The interpretation of single source and mixed DNA profiles, \textit{Forensic Science International: Genetics},  \textbf{7}, 516--28. \\\texttt{https://www.strmix.com/}
			\item  Balding, D. J.  (2013).  Evaluation of mixed-source, low-template DNA profiles in forensic science. \textit{Proceedings of the National Academy of Sciences of USA}. 
			 \\
			 \texttt{https://CRAN.R-project.org/package=likeLTD}.
		\item  Hernandis, E.,  D{\o}rum, G. and  Egeland, T. (2019). \texttt{relMix}: An open source software for DNA mixtures
		 with related contributors. \textit{Forensic Science International: Genetics Supplement Series}, \textbf{7}, 221--3.	
		\texttt{https://CRAN.R-project.org/package=relMix}.
			
		\item	 Haned, H. and  Gill, P. (2011). Analysis of complex DNA mixtures using the \texttt{Forensim} package, \textit{Forensic Science International: Genetics Supplement Series}, \textbf{3}, e79--e80. \\ 
		 \texttt{https://cran.r-project.org/web/packages/forensim/}
				
			%\item[8] \texttt{https://www.cybgen.com/products/casework.shtml}
		%\end{tablenotes}
			\end{enumerate}
	%\end{threeparttable}
%\end{table}


\begin{thebibliography}{}

\bibitem[\protect\citeAY{Balding}{2005}]{balding:05}
Balding, D.~J. (2005).
\newblock {\em Weight-of-Evidence for Forensic {DNA} Profiles}. Wiley, New
  York.

\bibitem[\protect\citeAY{Balding and Nichols}{1994}]{djb/ran:fsi}
Balding, D.~J. and Nichols, R.~A. (1994).
\newblock {DNA} profile match probability calculation: How to allow for
  population stratification, relatedness, database selection and single bands.
\newblock {\em Forensic Science International}, {\bf 64}, 125--40.

\bibitem[\protect\citeAY{Cotterman}{1940}]{cotterman}
Cotterman, C.~W. (1940).
\newblock {\em A calculus for statistico-genetics}.
\newblock PhD thesis, The Ohio State University.

\bibitem[\protect\citeAY{Cowell}{2016}]{cowell2016combining}
Cowell, R. (2016).
\newblock Combining allele frequency uncertainty and population substructure
  corrections in forensic {DNA} calculations.
\newblock {\em Forensic Science International: Genetics}, {\bf 23}, 210--6.

\bibitem[\protect\citeAY{Cowell {\it et~al}.}{2015}]{cowell:etal:15}
Cowell, R.~G., Graversen, T., Lauritzen, S., and Mortera, J. (2015).
\newblock Analysis of {DNA} mixtures with artefacts.
\newblock {\em Journal of the Royal Statistical Society Series C (with
  discussion)}, {\bf 64}, 1--48.

\bibitem[\protect\citeAY{Good}{1979}]{good}
Good, I.~J. (1979).
\newblock Studies in the {History of Probability and Statistics}. {XXXVII A. M.
  Turing's} statistical work in {World War II}.
\newblock {\em Biometrika}, {\bf 66}, 393--6.

\bibitem[\protect\citeAY{Graversen}{2013}]{graversen:package:13}
Graversen, T. (2013).
\newblock {\em {DNAmixtures}: Statistical Inference for Mixed Traces of DNA}.
\newblock R package version 0.1-0, dnamixtures.r-forge.r-project.org/.

\bibitem[\protect\citeAY{Graversen and
  Lauritzen}{2015}]{graversen:lauritzen:comp:13}
Graversen, T. and Lauritzen, S. (2015).
\newblock Computational aspects of {DNA} mixture analysis.
\newblock {\em Statistics and Computing}, {\bf 25}, 527--41.
\newblock arXiv:1302.4956.

\bibitem[\protect\citeAY{Green}{2015}]{green:cglmdiscn}
Green, P.~J. (2015).
\newblock Contribution to {D}iscussion of paper by {C}owell, et al.
\newblock {\em Journal of the Royal Statistical Society Series C}, {\bf 64},
  41.

\bibitem[\protect\citeAY{Green}{2020a}]{kinmix}
Green, P.~J. (2020a).
\newblock {\em KinMix: DNA mixture analysis with related contributors}.
\newblock R package 2.0, https://petergreenweb.wordpress.com/kinmix2-0.

\bibitem[\protect\citeAY{Green}{2020b}]{kinmix-userguide}
Green, P.~J. (2020b).
\newblock {\em KinMix User Guide 2.0: DNA mixture analysis with related
  contributors}.
\newblock https://petergreenweb.wordpress.com/kinmix2-0.

\bibitem[\protect\citeAY{Green and Mortera}{2009}]{green:mortera:09}
Green, P.~J. and Mortera, J. (2009).
\newblock Sensitivity of inferences in forensic genetics to assumptions about
  founder genes.
\newblock {\em Annals of Applied Statistics}, {\bf 3}, 731--63.

\bibitem[\protect\citeAY{Green and Mortera}{2017}]{green:mortera:17}
Green, P.~J. and Mortera, J. (2017).
\newblock Paternity testing and other inference about relationships from {DNA}
  mixtures.
\newblock {\em Forensic Science International: Genetics}, {\bf 28}, 128--37.

\bibitem[\protect\citeAY{Green {\it et~al}.}{2021}]{green-mortera-prieto}
Green, P.~J., Mortera, J., and Prieto, L. (2021).
\newblock Casework applications of probabilistic genotyping methods for {DNA}
  mixtures that allow relationships between contributors.
\newblock {\em Forensic Science International: Genetics}, {\bf 52}, 102482.

\bibitem[\protect\citeAY{Hansson}{2017}]{pcrsim}
Hansson, O. (2017).
\newblock {\em pcrsim: Simulation of the Forensic DNA Process}.
\newblock R package version 1.0.2, https://CRAN.R-project.org/package=pcrsim.

\bibitem[\protect\citeAY{Hernandis {\it et~al}.}{2019}]{hernandis2019relmix}
Hernandis, E., D{\o}rum, G., and Egeland, T. (2019).
\newblock relmix: An open source software for {DNA} mixtures with related
  contributors.
\newblock {\em Forensic Science International: Genetics Supplement Series},
  {\bf 7}, 221--3.

\bibitem[\protect\citeAY{Jacquard}{1974}]{jacquard}
Jacquard, A. (1974).
\newblock {\em The genetic structure of populations}. Springer-Verlag.

\bibitem[\protect\citeAY{Lauritzen and Spiegelhalter}{1988}]{laur/spieg}
Lauritzen, S.~L. and Spiegelhalter, D.~J. (1988).
\newblock Local computations with probabilities on graphical structures and
  their application to expert systems (with discussion).
\newblock {\em Journal of the Royal Statistical Society, Series B}, {\bf 50},
  157--224.

\bibitem[\protect\citeAY{Mortera}{2020}]{mortera-arsia}
Mortera, J. (2020).
\newblock {DNA} mixtures in forensic investigations: The statistical state of
  the art.
\newblock {\em Annu. Rev. Stat. Appl.}, {\bf 7}, 1--34.

\bibitem[\protect\citeAY{Nadot and Vaysseix}{1973}]{nadot:vaysseix}
Nadot, R. and Vaysseix, G. (1973).
\newblock Apparentement et identit\'e. {A}lgorithme du calcul des coefficients
  d'identit\'e.
\newblock {\em Biometrics}, {\bf 29}, 347--59.

\bibitem[\protect\citeAY{Thompson}{1974}]{Thompson74}
Thompson, E.~A. (1974).
\newblock Gene identities and multiple relationships.
\newblock {\em Biometrics}, {\bf 30}, 667--80.

\bibitem[\protect\citeAY{Thompson}{2013}]{thompson:genetics}
Thompson, E.~A. (2013).
\newblock Identity by descent: Variation in meiosis, across genomes, and in
  populations.
\newblock {\em Genetics}, {\bf 194}, 301--26.

\bibitem[\protect\citeAY{Tvedebrink}{2010}]{tvedebrink}
Tvedebrink, T. (2010).
\newblock Overdispersion in allelic counts and $\theta$-correction in forensic
  genetics.
\newblock {\em Theoretical Population Biology}, {\bf 78}, (3), 200 -- 210.

\bibitem[\protect\citeAY{Tvedebrink {\it et~al}.}{2015}]{tvedebrink:15}
Tvedebrink, T., Eriksen, P., and Morling, N. (2015).
\newblock The multivariate {D}irichlet-multinomial distribution and its
  application in forensic genetics to adjust for subpopulation effects using
  the $\theta$-correction.
\newblock {\em Theoretical Population Biology}, {\bf 105}, 24--32.

\bibitem[\protect\citeAY{Vigeland}{2019a}]{pedtools}
Vigeland, M.~D. (2019a).
\newblock {\em pedtools: Creating and Working with Pedigrees and Marker Data}.
\newblock R package version 0.9.0, https://github.com/magnusdv/pedtools.

\bibitem[\protect\citeAY{Vigeland}{2019b}]{ribd}
Vigeland, M.~D. (2019b).
\newblock {\em ribd: Pedigree-based Relatedness Coefficients}.
\newblock R package version 1.0.0, https://github.com/magnusdv/ribd.

\bibitem[\protect\citeAY{Wright}{1940}]{wright40}
Wright, S. (1940).
\newblock Breeding structure of populations in relation to speciation.
\newblock {\em American Naturalist}, {\bf 74}, 232--48.

\bibitem[\protect\citeAY{Wright}{1951}]{wright51}
Wright, S. (1951).
\newblock The genetical structure of populations.
\newblock {\em Annals of Eugenics}, {\bf 15}, 323--54.

\end{thebibliography}
\end{document}